\documentclass{article} 
\usepackage{iclr2027_conference,times}

\usepackage{amsmath,amsfonts,bm}

\def\eqref#1{equation~\ref{#1}}

\def\1{\bm{1}}

\DeclareMathAlphabet{\mathsfit}{\encodingdefault}{\sfdefault}{m}{sl}
\SetMathAlphabet{\mathsfit}{bold}{\encodingdefault}{\sfdefault}{bx}{n}

\DeclareMathOperator*{\argmin}{arg\,min}

\usepackage[bookmarks=true,breaklinks=true,colorlinks=true,linkcolor=Plum,citecolor=Plum]{hyperref}
\usepackage{url}
\usepackage[table,dvipsnames]{xcolor}
\usepackage{pifont}
\usepackage{times}
\usepackage{array}
\usepackage{booktabs}
\usepackage{tabularx}
\usepackage{multirow}
\usepackage{multicol}
\usepackage{wrapfig}
\usepackage{gensymb}
\usepackage{amsmath}
\usepackage{amsthm}
\usepackage{amssymb}
\usepackage{algorithm2e}
\usepackage{algorithmic}
\usepackage{graphicx}
\usepackage{subcaption}
\usepackage[normalem]{ulem}
\usepackage{ifthen}
\usepackage{enumerate}
\usepackage{enumitem}
\usepackage{listings}
\usepackage{textcomp}
\usepackage{makecell}
\usepackage{pdfpages}
\usepackage{cleveref}
\usepackage{soul}
\usepackage{cite}
\usepackage{tcolorbox}
\usepackage[compact]{titlesec}
\usepackage{pgfgantt}
\setlist{noitemsep, leftmargin=*, topsep=0pt, partopsep=0pt}

\lstdefinestyle{promptstyle}{basicstyle=\ttfamily,breaklines=true,frame=single}

\crefformat{section}{\S#2#1#3} 
\crefformat{subsection}{\S#2#1#3}
\crefformat{subsubsection}{\S#2#1#3}

\newcommand{\SYSTEM}{\texttt{PRISM}}

\newcommand*\circled[1]{\tikz[baseline=(char.base)]{\small{\textbf{
			\node[shape=circle,fill,inner sep=0.75pt] (char) {\textcolor{white}{#1}};}}}}

\newtheoremstyle{boldproposition}
  {3pt}                          
  {3pt}                          
  {\normalfont}                  
  {}                             
  {\bfseries}                    
  {.}                            
  {0.5em}                        
  {\thmname{#1}\thmnumber{ #2}\thmnote{ \textbf{(#3)}}}

\theoremstyle{boldproposition}
\newtheorem{proposition}{Proposition}
\crefname{proposition}{Proposition}{Propositions}
\Crefname{proposition}{Proposition}{Propositions}

\title{Beyond Energy: When Sustainability Dimensions Reshape LLM Serving Decisions}

\author{Tianyao Shi, Xipeng Shen, Yi Ding  \\ 
Elmore Family School of Electrical and Computer Engineering, Purdue University, USA \\
\texttt{\{shi676,shen810,yiding\}@purdue.edu} 
}

\iclrfinalcopy 
\begin{document}

\maketitle

\begin{abstract}

Large language model (LLM) serving has environmental impacts across energy consumption, carbon emission, water consumption, and biodiversity loss. Yet these dimensions are largely evaluated in isolation, leaving it unclear when and how they lead to different optimization decisions. We present \SYSTEM{}, a unified framework for characterizing and optimizing LLM serving across energy, carbon, water, and biodiversity impacts. Our analysis reveals a fundamental distinction: computing configurations determine energy consumption, whereas where and when LLM serving is deployed determine its carbon, water, and biodiversity impacts. Under a fixed deployment choice and operational-only accounting, all dimensions preserve the same energy-based configuration ranking. Deployment rankings can diverge across dimensions, while embodied impacts can break configuration invariance when they exceed a lifecycle crossover boundary. \SYSTEM{} identifies these conditions, quantifies cross-dimensional regrets, and balances the four dimensions. In regional-routing experiments, \SYSTEM{} reduces median worst-case regret by 50.2\% relative to the strongest baseline.

\end{abstract}
\section{Introduction}

The rapid adoption of large language models (LLMs) has raised growing concerns about their environmental impact~\citep{lambert2026cradle,chien2026strategies,ding2024sustainable}. These concerns initially centered on the substantial \textbf{energy consumption}~\citep{strubell2019energy,fernandez2025energy,patel2024characterizing,stojkovic2025dynamollm} of LLM serving and later expanded to the associated \textbf{carbon emissions}~\citep{patterson2021carbon,luccioni2023estimating}. Recent work further highlighted the substantial \textbf{water consumption}~\citep{Ali2024making,wu2025not} from LLM systems through datacenter cooling, electricity generation, and hardware manufacturing. Their lifecycle activities, such as resource extraction and land use, can also contribute to \textbf{biodiversity loss}~\citep{shi2025servers,shi2026birds}. Together, energy, carbon, water, and biodiversity capture distinct aspects of LLM serving sustainability~\citep{shi2026sustainability}.

Most existing sustainable AI research nevertheless evaluates LLM systems through a single environmental dimension. However, the four dimensions depend on different factors. Energy depends on workloads, models, hardware, and serving configurations~\citep{chung2026ml,shi2025systematic}; carbon additionally depends on the electricity mix and embodied emissions~\citep{nguyen2024towards,li2024towards}; water depends on cooling types, electricity-generation technologies, manufacturing, and local water scarcity~\citep{wu2025not,jiang2025thirstyflops}; and biodiversity aggregates multiple pathway-specific effects on ecosystems~\citep{shi2025servers,shi2026birds}. Although interconnected, these dimensions are neither equivalent nor necessarily proportional.

Beyond studying these dimensions separately, existing work provides limited insight into how they affect decisions. Prior sustainability-aware cloud systems demonstrate carbon--water trade-offs~\citep{jegham2025hungry,jiang2025waterwise} and the benefits of regional workload shifting~\citep{gsteiger2024caribou}, but typically consider only one or two dimensions, target general cloud workloads, and do not explain when or why computing-configuration and deployment rankings agree or diverge for LLM serving workloads. We therefore ask two central questions: \circled{1} When do energy, carbon, water, and biodiversity agree on computing configurations and deployment choices, and what causes their rankings to diverge? \circled{2} When they disagree, how should an LLM serving system balance the four dimensions while satisfying performance and quality requirements?

We present \SYSTEM{}, a unified framework for characterizing and optimizing LLM serving across energy, carbon, water, and biodiversity. \SYSTEM{} evaluates the four dimensions using consistent system boundaries and functional units across workloads, computing configurations, and deployment choices. It analyzes configuration and deployment rankings, identifies lifecycle crossover conditions, quantifies the disagreement through cross-dimensional regret, and optimizes deployment choices under multi-dimensional objectives. This paper makes the following contributions:
\begin{itemize}
    \item We develop \SYSTEM{}, a unified framework for characterizing and optimizing energy, carbon, water, and biodiversity impacts of LLM serving under consistent system boundaries and functional units.
    \item We establish operational configuration invariance: under a fixed deployment choice and operational-only accounting, all four dimensions preserve the configuration ranking induced by IT energy. We further derive a lifecycle crossover condition that identifies when    configuration-dependent embodied impacts break this invariance.
    \item We systematically characterize the four dimensions across workloads, computing configurations, deployment choices, and quantify when their rankings agree or diverge.
    \item We formulate multidimensional deployment optimization that balances the four dimensions by limiting the largest relative penalty without directly combining impacts.
\end{itemize}

Our results reveal a fundamental distinction between computing configurations and deployment choices. Under a fixed deployment choice, computing configurations change impact magnitude but preserve the same operational configuration ranking across dimensions; embodied impacts change
the minimum-impact configuration in only 4 of 882 evaluated settings. In contrast, deployment rankings exhibit stronger disagreement: among six representative regions, selecting the energy-minimizing region incurs 194\% higher carbon emissions, 4,637\% higher water impact, and 195\% higher biodiversity impact than minimizing each respective dimension. When balancing all four dimensions, \SYSTEM{} achieves the lowest median worst-case regret among the evaluated methods, with a 50.2\% median reduction relative to the strongest baseline. These findings show that energy can guide configuration selection under fixed deployment and operational-only accounting, but cannot serve as a reliable proxy for other impacts when selecting where and when to deploy LLM serving.

\section{Multidimensional Impact Model and Decision Properties}
\label{sec:sustainability_dimensions}

We first unify established accounting methods for energy ($E$), carbon ($C$), water ($W$), and biodiversity ($B$) under a common formulation and then derive their implications for LLM serving. Detailed accounting methods for each dimension are provided in Appendix~\ref{app:impact_accounting}.

Let $w$ denote the \emph{workload}, including the requests or tasks to be served and the functional unit~\citep{wu2025unveiling} used for comparison. We vary workloads to study how their characteristics affect environmental impacts. Let $x$ denote a \emph{computing configuration}, including the model, GPU type, GPU count, and parallelism. A \emph{deployment choice} $(r,t)$ specifies the region $r$ and execution time period $t$. We distinguish between \textbf{operational}, \textbf{embodied}, and \textbf{lifecycle} impacts~\citep{gupta2022act}. Operational impact arises from energy consumed while running the workload. Embodied impact arises from manufacturing the hardware and is allocated to the workload based on its use of that hardware. Within our system boundary, lifecycle impact refers to the sum of operational and embodied impacts. Our lifecycle system boundary~\citep{suh2004system} includes hardware manufacturing and system operation; transportation and end-of-life are excluded because detailed inventory data are unavailable. We first measure the IT energy consumed by workload $w$ under configuration $x$, denoted by $E_{\mathrm{IT}}(x,w)$. The total operational energy attributed to the workload is
\begin{equation}
E_{\mathrm{op}}(x,r,t,w)
=
E_{\mathrm{IT}}(x,w)\cdot \mathrm{PUE}(r,t),
\label{eq:operational_energy}
\end{equation}
where $\mathrm{PUE}(r,t)$ denotes the Power Usage Effectiveness (PUE) of a datacenter in region $r$ during time $t$, defined as the ratio of total facility energy consumption to IT equipment energy consumption~\citep{barroso2019datacenter}. Although the four sustainability dimensions characterize different outcomes, their operational components share a common structure:
\begin{equation}
I_{m,\mathrm{op}}(x,r,t,w)
=
E_{\mathrm{IT}}(x,w)\cdot \alpha_m(r,t),
\qquad
m\!\in\!\{E,C,W,B\},
\label{eq:operational_common_form}
\end{equation}
where $\alpha_m(r,t)$ is the \emph{operational impact intensity} of dimension $m$, defined as the operational impact produced per unit of IT energy. It is derived from three types of regional and temporal data: facility factors, such as PUE and Water Usage Effectiveness (WUE)~\citep{Ali2024making}; electricity-system environmental intensities, such as carbon intensity (CI)~\citep{maji2022carboncast,yan2025ensembleci}, electricity water intensity (EWIF)~\citep{wu2025not}, and grid biodiversity intensity ($\mathrm{BIF}$)~\citep{shi2025servers}; and local characterization factors. Specifically, water stress factor (WSF)~\citep{wu2025not} adjusts water consumption for local water stress, while $\mathrm{CF}_{B,W}$ converts direct local water consumption into biodiversity impact~\citep{shi2026birds}. We refer to these inputs collectively as \emph{environmental data}. The LLM serving workload and computing configuration determine $E_{\mathrm{IT}}(x,w)$, while deployment choice determines $\alpha_m(r,t)$. Adding the embodied component gives the lifecycle impact
\begin{equation}
I_m(x,r,t,w)
=
E_{\mathrm{IT}}(x,w)\cdot\alpha_m(r,t)
+
I_{m,\mathrm{emb}}(x,w).
\label{eq:unified_lifecycle_model}
\end{equation}
\Cref{tab:sustainability_dimensions} summarizes how the four dimensions instantiate this common model. The complete derivations of each intensity, factor, and embodied component are in Appendix~\ref{app:impact_accounting}. 

Within dimension $m$, decisions are ranked by increasing impact. A \emph{configuration ranking} compares $x$ while holding $(w,r,t)$ fixed, whereas a \emph{deployment ranking} compares $(r,t)$ while holding $(x,w)$ fixed. Two dimensions disagree when they reverse the ordering of at least one pair of configurations or deployment choices.

\begin{table*}[t]
\centering
\caption{Unified operational and embodied accounting across four sustainability dimensions.}
\vspace{-0.1in}
\label{tab:sustainability_dimensions}
\resizebox{\textwidth}{!}{
\begin{tabular}{llll}
\toprule
\textbf{Dimension}
&
\textbf{Effective Operational Intensity $\alpha_m(r,t)$}
&
\textbf{Embodied Component}
&
\textbf{Output}
\\
\midrule
Energy
&
$\mathrm{PUE}(r,t)$
&
---
&
kWh
\\
Carbon
&
$\mathrm{PUE}(r,t)\cdot\mathrm{CI}(r,t)$
&
$C_{\mathrm{emb}}(x,w)$
&
kg CO$_2$e
\\
Water
&
$[\mathrm{WUE}(r,t)+\mathrm{PUE}(r,t)\cdot\mathrm{EWIF}(r,t)]\cdot\mathrm{WSF}(r)$
&
$W_{\mathrm{emb}}(x,w)$
&
m$^3$ world-eq
\\
Biodiversity
&
$\mathrm{PUE}(r,t)\cdot\mathrm{BIF}(r,t)+\mathrm{WUE}(r,t)\cdot\mathrm{WSF}(r)
\cdot\mathrm{CF}_{B,W}(r)$
&
$B_{\mathrm{emb}}(x,w)$
&
species$\cdot$year
\\ 
\bottomrule
\end{tabular}
}
\vspace{-0.2in}
\end{table*}

\begin{proposition}[Operational Configuration Invariance]
\label{prop:configuration_invariance}
For a fixed deployment choice, operational sustainability dimensions preserve the configuration ranking induced by IT energy:
\begin{equation}
E_{\mathrm{IT}}(x_i,w)<E_{\mathrm{IT}}(x_j,w)
\Longleftrightarrow
I_{m,\mathrm{op}}(x_i,r,t,w)
<
I_{m,\mathrm{op}}(x_j,r,t,w),
\label{eq:configuration_invariance}
\end{equation}
where $x_i$ and $x_j$ are two computing configurations. This follows because $\alpha_m(r,t)$ is positive and constant across computing configurations within the same deployment choice. 
\end{proposition}

\begin{proposition}[Operational Deployment Ranking Invariance]
\label{prop:regional_invariance}
For a fixed workload and sustainability dimension, the operational deployment ranking is independent of LLM computing configurations. For two choices $(r_i,t_i)$ and $(r_j,t_j)$,
\begin{equation}
I_{m,\mathrm{op}}(x,r_i,t_i,w)
<
I_{m,\mathrm{op}}(x,r_j,t_j,w)
\Longleftrightarrow
\alpha_m(r_i,t_i)
<
\alpha_m(r_j,t_j).
\label{eq:regional_invariance}
\end{equation}
The common positive factor $E_{\mathrm{IT}}(x,w)$ cancels when comparing regions. Changing LLM computing configuration therefore changes the magnitude of operational impact but not the deployment ranking. 
\end{proposition}

\begin{proposition}[Lifecycle Crossover]
\label{prop:lifecycle_crossover}
Configuration-dependent embodied impacts can break operational configuration invariance. Consider configurations $x_i$ and $x_j$ such that
$E_{\mathrm{IT}}(x_i,w)<E_{\mathrm{IT}}(x_j,w)$. Dimension $m$ prefers the less energy-efficient configuration $x_j$ when
\begin{equation}
I_{m,\mathrm{emb}}(x_i,w)
-
I_{m,\mathrm{emb}}(x_j,w)
>
\alpha_m(r,t)
\left[
E_{\mathrm{IT}}(x_j,w)
-
E_{\mathrm{IT}}(x_i,w)
\right].
\label{eq:lifecycle_crossover}
\end{equation}
This condition defines the crossover boundary at which the embodied-impact advantage of $x_j$ exceeds its operational disadvantage compared to $x_i$.
\end{proposition}
\section{\SYSTEM{}: Unified Characterization and Decision Optimization}
\label{sec:framework}

\begin{figure*}[t]
\centering
\includegraphics[width=\textwidth]{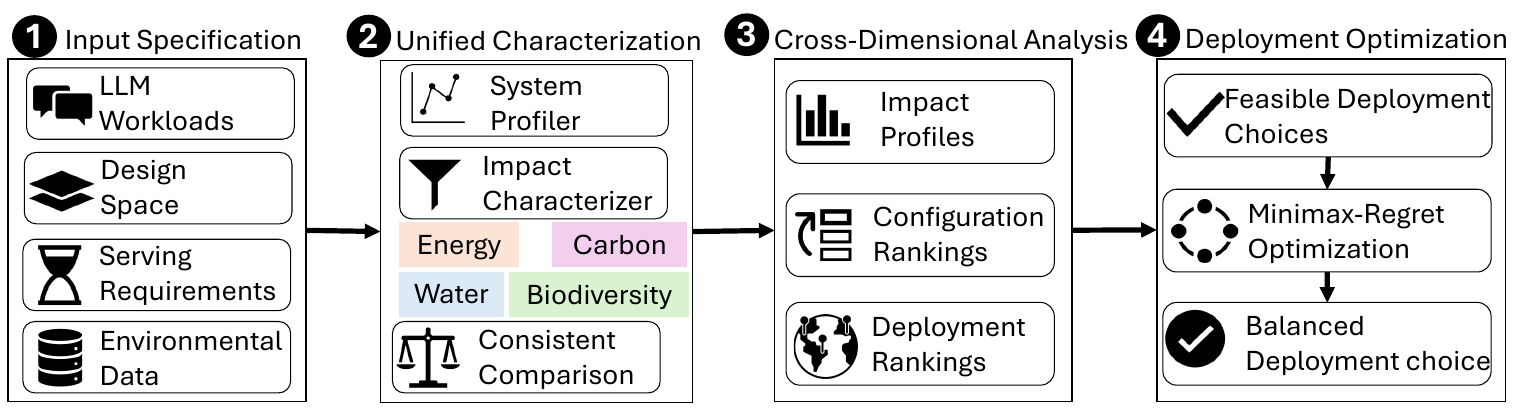}
\vspace{-0.2in}
\caption{Overview of \SYSTEM{}.}
\label{fig:overview}
\vspace{-0.2in}
\end{figure*}

\Cref{fig:overview} presents \SYSTEM{}, our framework for characterizing LLM serving and optimizing its deployment across energy, carbon, water, and biodiversity. \SYSTEM{} connects four stages: input specification, unified characterization, cross-dimensional analysis, and deployment optimization. Additional details on \SYSTEM{}'s implementation and decision analysis are provided in Appendix~\ref{app:prism_details}. 

\textbf{Input Specification and Unified Characterization.}
\SYSTEM{} takes as input a workload context $w$, design space, deployment choices, service requirements, and the regional and lifecycle data required for environmental characterization. \SYSTEM{} profiles each computing configuration to measure IT energy, latency, throughput, and output quality or task success, retaining only configurations that satisfy the same service requirements. It then combines the measured system profile with environmental data to calculate energy, carbon, water, and biodiversity impacts using models in~\Cref{sec:sustainability_dimensions}. All dimensions use the same workload execution, system boundary, and functional unit, with impacts reported per request or per successfully completed task. Let $d=(x,r,t)$ denote a serving decision. For $m\!\in\!\{E,C,W,B\}$, $I_m(d,w)\!\equiv\!I_m(x,r,t,w)$ denotes the impact of executing $w$ under $d$.

\textbf{Cross-Dimensional Analysis.}
\SYSTEM{} analyzes the four sustainability dimensions along two decision axes: computing configuration and deployment choice. For a fixed deployment choice, it compares configuration rankings across four dimensions and evaluates the operational configuration-invariance property in~\Cref{prop:configuration_invariance}. For lifecycle analysis, it applies~\Cref{prop:lifecycle_crossover} to calculate the embodied-impact difference required to reverse each observed operational ranking. For a fixed computing configuration, it compares deployment rankings to determine when the four dimensions lead to different deployment choices. To quantify deployment disagreement, let $d_i^*$ denote the deployment minimizing dimension $i$. The regret incurred under dimension $j$ when selecting $d_i^*$ is
\begin{equation}
R_{i\rightarrow j}(w)
=
\frac{
I_j(d_i^*,w)-I_j(d_j^*,w)
}{
I_j(d_j^*,w)
}.
\label{eq:cross_dimensional_regret}
\end{equation}
A small $R_{i\rightarrow j}$ indicates that optimizing dimension $i$ produces a deployment choice close to the optimum for dimension $j$. 

\textbf{Deployment Optimization.}
According to~\Cref{prop:configuration_invariance}, all four dimensions preserve the energy-based configuration ranking under the same deployment choice. \SYSTEM{} first selects the minimum-energy feasible configuration $x_E^*$ for each workload. Let $s\in\mathcal{S}$ denote a deployment choice, where $\mathcal{S}$ contains all feasible deployment choices subject to service and capacity constraints. For a single workload, $s$ reduces to one serving decision $d=(x_E^*,r,t)$; for a workload trace, it assigns each request $k$ to a region and execution time period. The aggregate impact of a deployment choice is
\begin{equation}
I_m(s)
=
\sum_k I_m(x_E^*(w_k),r_k,t_k,w_k),
\qquad
I_m^*=\min_{s\in\mathcal{S}} I_m(s).
\end{equation}
For multi-dimensional optimization, \SYSTEM{} identifies the Pareto-efficient deployment and selects
\begin{equation}
s_{\SYSTEM{}}^*
=
\mathop{\mathrm{arg\,min}}_{s\in\mathcal{S}}
\max_{m\in\{E,C,W,B\}}
\frac{I_m(s)-I_m^*}{I_m^*}.
\label{eq:multidimensional_deployment_general}
\end{equation}
This formulation minimizes the largest relative loss from the dimension-specific optimum, allowing the four dimensions to be balanced without aggregating into one single objective.

\section{Evaluation Methodology}
\label{sec:evaluation_methodology}

We evaluate \SYSTEM{} through three research questions (RQs). \textbf{RQ1 (Characterization):} How do LLM workloads, configurations, and deployment choices affect the magnitude and variation of energy, carbon, water, and biodiversity impacts? \textbf{RQ2 (Ranking Disagreement)}: When do sustainability dimensions lead to different configuration or deployment rankings, and what causes these difference? \textbf{RQ3 (Multi-dimensional Optimization):} How effectively does \SYSTEM{} balance the four dimensions while satisfying performance and quality requirements?

\textbf{Setup.}
We evaluate both non-agentic and agentic LLM serving workloads. ShareGPT~\citep{sharegpt} represents interactive conversations, LongBench~\citep{bai2023longbench} represents long-context tasks, RepoBench~\citep{liu2023repobench} represents IDE-level code completion tasks, and SWE-bench Verified~\citep{jimenez2024swebench} represents agentic software-engineering tasks. Our design space includes models from the Llama-3~\citep{grattafiori2024llama}, GPT-OSS~\citep{openai2025gptoss120bgptoss20bmodel}, Qwen3~\citep{qwen3technicalreport}, and Gemma-4~\citep{gemma4} families, deployed on NVIDIA A100, L40, and H100 GPUs. We vary the model, GPU, parallelism, and request load while retaining only configurations that satisfy performance and quality requirements. GPU energy is measured using NVML~\citep{nvml}, and non-GPU host energy is estimated from component utilization. We report impacts per request for non-agentic workloads and per successfully completed task for agentic workloads. We evaluate deployment across 46 locations (\Cref{tab:modelled-deployment-regions}) worldwide based on major cloud providers' official region documentation to span contrasting facility efficiency, electricity mixes, carbon and water intensities, water stress, and ecological conditions.

\textbf{Key Assumptions.}
We compare computing configurations and deployment choices that deliver workload outcomes under the same functional unit and service requirements. We assume that every evaluated computing configuration is available in all deployment choices and that a fixed computing configuration has the same IT-level execution profile across regions; regional conditions affect its impact through operational impact intensities and environmental data. Our optimization focuses on operational impacts because embodied impacts are already incurred for an existing hardware fleet and cannot be changed through workload scheduling or regional placement; including them in the optimization can therefore increase, rather than reduce, total environmental impact~\citep{bashir2024sunk,gsteiger2024caribou}. Complete workload statistics, model and hardware specifications, profiling and quality-evaluation protocols, quality scores, latency constraints, and regional data are provided in Appendix~\ref{app:setup}.

\section{Characterizing Four Dimensions}
\label{sec:characterization}

We first examine how workloads, computing configurations, and deployment choice affect energy, carbon, water, and biodiversity impacts. Beyond comparing their impact magnitudes, we ask whether the four dimensions rank the same computing configurations consistently.

\begin{figure}[t]
    \centering
    \includegraphics[width=\linewidth]{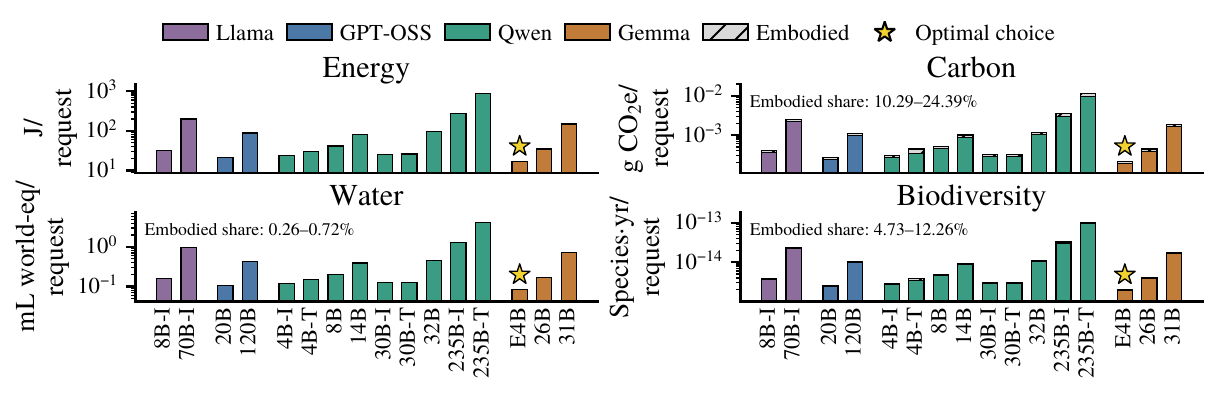}
    \vspace{-0.2in}
    \caption{Impact of model family and size on per-request energy, carbon, water, and biodiversity. Solid and hatched bars denote operational and embodied impacts, respectively; stars mark the minimum-impact configuration for each dimension.}
    \label{fig:main_charac}
    \vspace{-0.2in}
\end{figure}

\textbf{Configurations and Workloads.}
\Cref{fig:main_charac} provides a detailed characterization across model families and scales. We fix the workload to ShareGPT, use H100 GPUs, select the most energy-efficient tensor parallelism (TP) for each model, and apply France's 2024 annual-average environmental intensities. Impact generally increases with model size because larger models require more serving energy per request. Model size alone, however, does not determine impact: mixture-of-experts models can incur substantially lower impact than similarly sized dense models because they activate only a subset of their parameters per token. Despite measuring different sustainability outcomes, the four dimensions exhibit nearly identical trends and select the same minimum-impact model in this setting. Operational impact dominates carbon, water, and biodiversity, and their values therefore largely scale with the same underlying serving energy. The allocated embodied components change their magnitudes but are insufficient to reverse the preferred configuration in this example.

\begin{figure}[t]
    \centering
    \includegraphics[width=\linewidth]{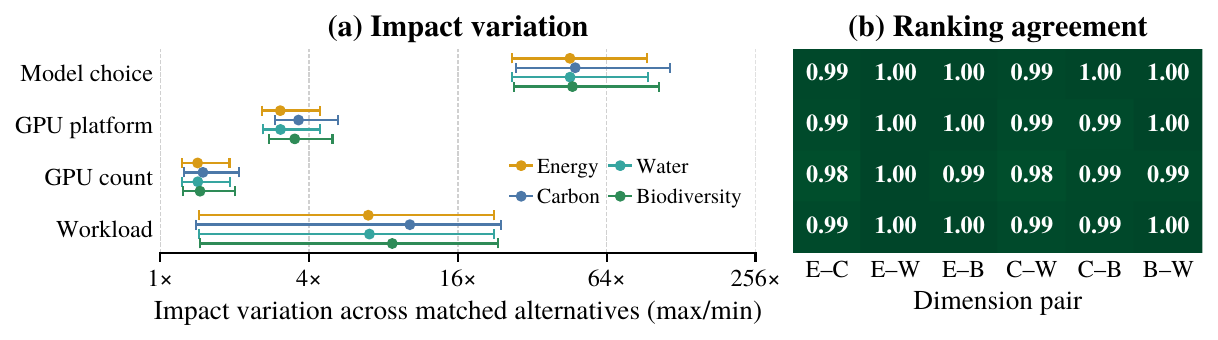}
    \vspace{-0.2in}
    \caption{Summary of computing configuration and workload effects under a fixed deployment choice. (a) Median and interquartile range of impact variation when varying model choice, GPU platform, GPU count, or non-agentic workload. (b) Cross-dimensional rank agreement for the same comparisons. Each comparison varies only the indicated category and uses a same functional unit.}
    \vspace{-0.2in}
    \label{fig:configuration_summary}
\end{figure}

\Cref{fig:configuration_summary} summarizes whether this pattern generalizes across model and scale, GPU platform, TP, and workload types. These factors substantially change impact magnitude. H100 generally achieves lower per-request impact than A100 and L40, whereas the effect of TP depends on scaling efficiency: additional GPUs reduce impact only when their throughput improvement offsets the added energy consumption. Workload properties also matter. The same model and hardware configuration show a large variance of per-request impact among workloads with distinct prompt and generation length distributions (\Cref{tab:serving_workload_length_stats}), resulting in up to 23.8$\times$ ratio between the shortest chat conversations and the longest document summarizations. Agentic workloads are excluded from this comparison because their impacts are measured using a different functional unit: impact per successfully completed task. Appendix~\ref{app:add_res_workload} shows that SWE-Verified incurs much larger per-task impact due to repeated model invocations and long execution, while lower task success can cause a smaller model to have higher impact per successful task by amortizing failed attempts over fewer successes.

Nevertheless, under fixed deployment choices, carbon, water, and biodiversity impacts remain strongly aligned with the energy-based ranking across computing configurations. Computing configurations and workloads determine how much impact is produced, but the deployment choice among sustainability dimensions usually does not change which configuration is preferred. Detailed results for models, individual GPUs, TP levels, and workloads are provided in Appendix~\ref{app:add_res_hardware} and~\ref{app:add_res_workload}.

\textbf{Takeaway 1:}
\textit{Computing configurations substantially change the magnitudes of impacts in all four dimensions, but under fixed deployment choices, operational impacts of carbon, water, and biodiversity generally preserve the energy-based configuration ranking.}

\begin{figure}[t]
    \centering
    \includegraphics[width=\linewidth]{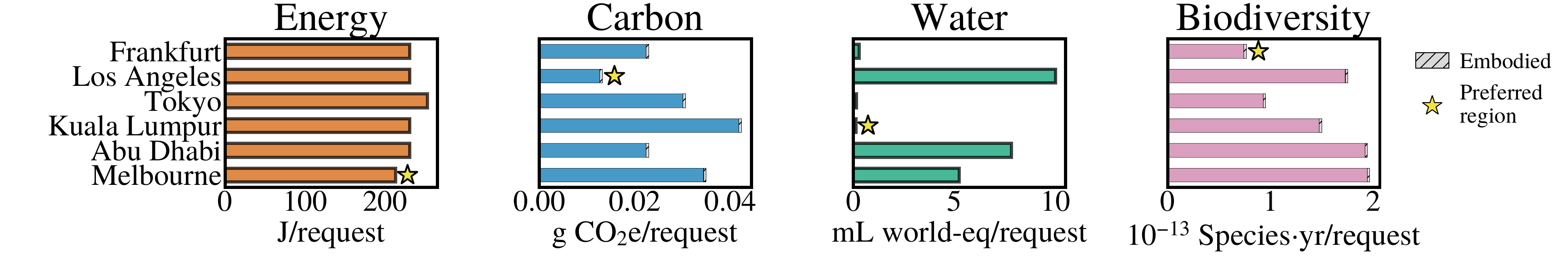}
    \vspace{-0.2in}
    \caption{Regional variation in per-request sustainability impacts for Qwen3-235B-A22B-Instruct on H100 (TP8). Stars mark the lowest-impact region for each dimension among the locations shown.}
    \label{fig:regional_comparison_simple}
    \vspace{-0.2in}
\end{figure}

\textbf{Region and Time.}
\Cref{fig:regional_comparison_simple} fixes the workload and computing configuration while varying the deployment region. We show six representative locations from our 46-location dataset using 2024 annual-average operational impact intensities. Energy changes the modestly because the IT-level execution is fixed and regional variation enters primarily through PUE. Carbon, water, and biodiversity impacts vary much more because they additionally depend on the electricity mix, cooling conditions, water stress, and operational impact intensities. These regional factors also produce different preferences. Among the locations shown, Melbourne minimizes energy, Los Angeles carbon, Kuala Lumpur water, and Frankfurt biodiversity. Thus, a region with favorable facility efficiency does not necessarily minimize its broader environmental impacts. The computing configuration determines the underlying IT energy demand, whereas the deployment choice determines how that demand translates into carbon, water, and biodiversity impacts. Regional preferences are also time-dependent. Monthly and hourly measurements (\Cref{fig:main_regional_time_sens}) exhibit ranking crossovers, indicating that a region preferred under annual-average conditions may not remain preferable at finer temporal resolutions. Detailed temporal results and the corresponding operational impact intensity trends are provided in Appendix~\ref{app:add_region_time}.

\noindent\textbf{Takeaway 2:}
\textit{Unlike configuration rankings, which are aligned across dimensions under operational accounting, deployment rankings can differ across dimensions, causing energy, carbon, water, and biodiversity to prefer different regions or execution times.}
\section{When Sustainability Dimensions Disagree}
\label{sec:decision_disagreement}

The preceding results show that, under a fixed deployment choice, operational impacts in all four dimensions preserve the configuration ranking induced by IT energy, but their deployment rankings can differ. We next examine two sources of cross-dimensional ranking disagreement. First, configuration-dependent embodied impacts can break operational configuration invariance when the lifecycle crossover condition in~\Cref{prop:lifecycle_crossover} is satisfied. Second, regional and temporal variation in operational impact intensities can cause the dimensions to produce different deployment rankings.

\begin{figure}[t]
    \centering
    \includegraphics[width=\linewidth]{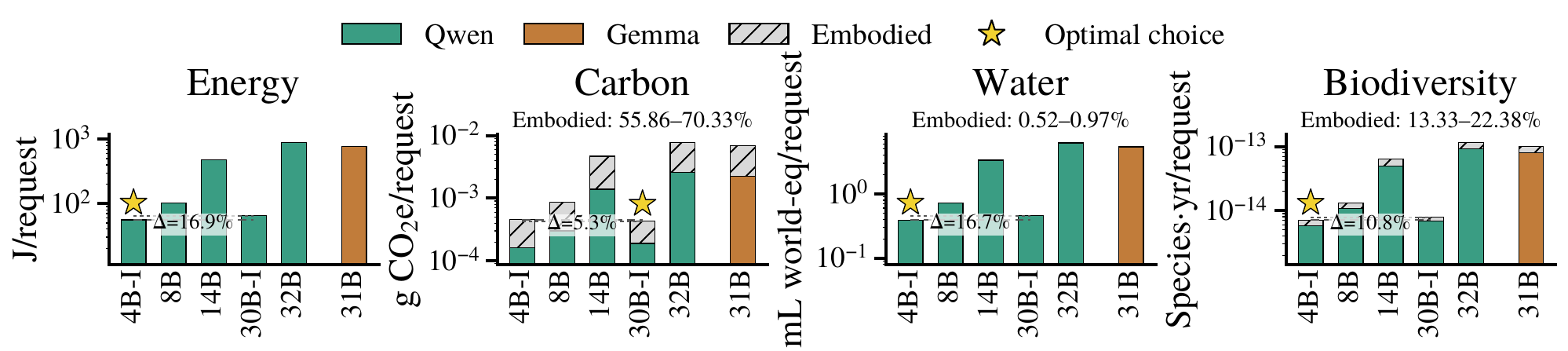}
    \vspace{-0.2in}
    \caption{Lifecycle-induced configuration ranking disagreement for RepoBench in Norway with edit-similarity $\geq44.65$. Stars mark the minimum-impact configuration for each dimension.}
    \label{fig:main_optimal_flip}
    \vspace{-0.2in}
\end{figure}

\textbf{Lifecycle-Induced Configuration Disagreement.}
Including configuration-dependent embodied impacts can break operational configuration invariance. \Cref{fig:main_optimal_flip} shows the largest disagreement among minimum-impact configurations observed in our quality-constrained evaluation. Energy, water, and biodiversity rank Qwen3-4B-Instruct on one H100 with TP1 first, whereascarbon ranks Qwen3-30B-Instruct on two H100s with TP2 first. The latter consumes 16.9\% more energy per request and increases water and biodiversity impacts by 16.7\% and 10.8\%, respectively, but produces lower carbon emissions. This carbon ranking reversal occurs because Norway's low carbon intensity makes the carbon impact from the additional energy consumption relatively small. Meanwhile, the higher throughput of Qwen3-30B amortizes its embodied carbon emission over more requests. For this configuration pair, the embodied-carbon advantage of Qwen3-30B is $1.83\times$ its additional operational-carbon penalty and therefore exceeds the lifecycle crossover boundary.

More generally, lifecycle configuration rankings can diverge only when the embodied-impact difference favors the higher-energy configuration and is large enough to exceed its region-specific operational penalty. The embodied share of total impact alone therefore does not predict a ranking reversal. Configuration ranking disagreement is rare in our evaluation: the dimensions select different minimum-impact configurations in only 4 of 882 quality- and performance-constrained settings (0.45\%). Appendix~\ref{app:lifecycle_disagreement} reports the remaining cases, controlled comparisons of GPU and parallelism choices, complete crossover calculations, and sensitivity to the assumed hardware lifetime.

\textbf{Takeaway 3:}
\textit{Lifecycle configuration rankings diverge only when the embodied-impact advantage of a higher-energy configuration exceeds its region-specific operational penalty. A large embodied share alone does not imply a ranking reversal.}

\begin{figure}[t]
    \centering
    \includegraphics[width=\linewidth]{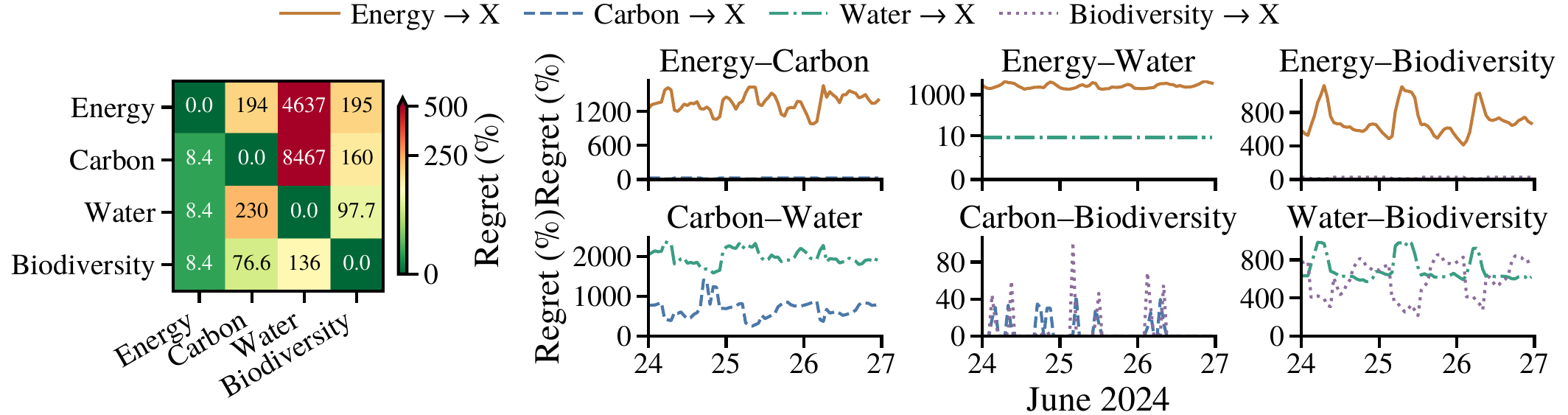}
    \vspace{-0.2in}
    \caption{Left: directional cross-dimensional regret among the dimension-optimal regions in \Cref{fig:regional_comparison_simple}. Cell $(i,j)$ reports the additional impact under dimension $j$ when selecting the region optimized for dimension $i$. Right: hourly directional cross-regret across selected regions during
    June 24--26, 2024, using the spatiotemporal operational impact intensities in
    \Cref{fig:main_regional_time_sens}.}
    \label{fig:regional_cross_regret}
    \vspace{-0.2in}
\end{figure}

\textbf{Regional and Temporal Deployment Ranking Disagreement.}
The four dimensions can produce different deployment rankings even under operational-only accounting because their operational impact intensities vary across regions. \Cref{fig:regional_cross_regret} (left) quantifies this disagreement among the six representative regions using directional cross-dimensional regret. Relative to the minimum-impact region under each dimension, choosing the energy-minimizing region increases carbon emissions by 194\%, water impact by 4,637\%, and biodiversity impact by 195\%. Minimizing operational energy is therefore not a reliable proxy for minimizing carbon, water, or biodiversity impacts when selecting a deployment region. These penalties are also strongly asymmetric. Although the energy-minimizing region performs substantially worse under the other three dimensions, the regions ranked first by carbon, water, and biodiversity increase operational energy by no more than 8.4\%. Thus, a modest increase in energy can correspond to a much larger reduction in another environmental impact.

Execution time also affects the consequences of deployment ranking disagreement. \Cref{fig:regional_cross_regret} (right) shows that directional regret changes considerably across hours as operational impact intensities vary. Some pairs, particularly energy and water, exhibit consistently high regret throughout the evaluated period, whereas the penalties between other pairs depend more strongly on execution time. Temporal scheduling can therefore reduce or increase the penalty of following one dimension's deployment ranking over another, but does not necessarily eliminate the underlying disagreement.

\noindent\textbf{Takeaway 4:}
\textit{Under operational-only accounting, cross-dimensional disagreement arises primarily in deployment rankings. Following the energy ranking can incur large and asymmetric carbon, water, and biodiversity penalties, while execution time changes the severity of these penalties but does not necessarily eliminate them.}

\section{Can \SYSTEM{} Balance Multiple Sustainability Dimensions?}

We evaluate how effectively \SYSTEM{} balances all four dimensions. We first instantiate \Cref{eq:multidimensional_deployment_general} as an offline regional-routing problem in which workload demand and hourly environmental data are known in advance. Its solution provides an oracle benchmark for evaluating rolling-horizon~\citep{sethi1991theory} online routing with limited future information. We use the first 24 hours of the Azure LLM Inference Dataset 2024~\citep{stojkovic2025dynamollm}, combining code and conversation requests with equal offered GPU demand. Each request is assigned within its arrival hour to one of the twelve regions in~\Cref{fig:main_regional_time_sens}, subject to regional capacity constraints. We use hourly operational impact intensities and generate heterogeneous regional capacities across 20 random seeds. We compare environment-unaware geographical load balancing; routing that individually minimizes energy, carbon, water, or biodiversity; an offline adaptation of WaterWise~\citep{jiang2025waterwise} that jointly optimizes carbon and water; and \SYSTEM{} that minimizes the largest normalized regret across four dimensions. We further evaluate rolling-horizon \SYSTEM{} and extend it to jointly select computing configurations and deployment regions for an agentic workload with completion time included as an additional objective. Trace construction, capacity generation, optimization formulations, baseline implementations, solver runtime, and sensitivity analyses are described in~\Cref{app:optimization_details}.

\begin{figure}[t]
    \centering
    \includegraphics[width=\linewidth]{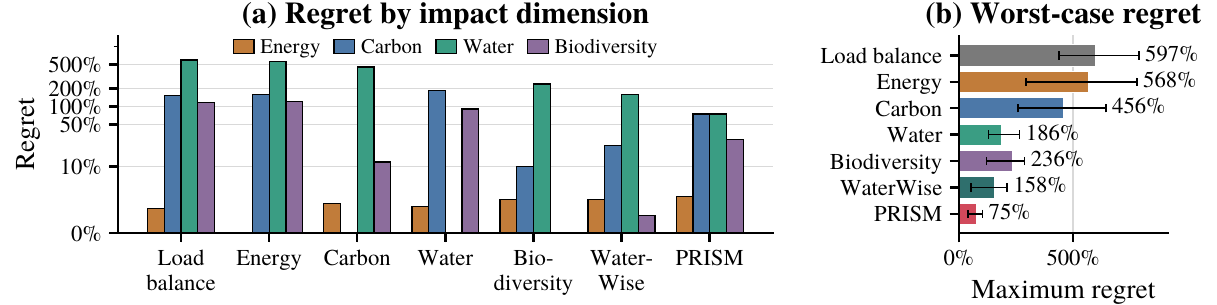}
    \vspace{-0.2in}
    \caption{Environmental trade-offs in offline regional routing. Regret measures the relative increase over the best feasible value for each impact. (a) Median regret for each dimension across 20 regional-capacity seeds; the symlog axis is linear below 10\%. (b) Median worst-case regret, with whiskers showing the full range across seeds. Lower is better. }
    \vspace{-0.2in}
    \label{fig:optimization_offline}
\end{figure}

\textbf{Balancing Four Dimensions.}
\Cref{fig:optimization_offline} shows that minimizing one dimension can impose a large penalty on another. Median worst-case regret ranges from 185.5\% for water-only routing to 568.0\% for energy-only routing, while environment-unaware load balancing reaches 597.2\%. WaterWise lowers median worst-case regret to 157.8\%, but its largest remaining penalty is its water regret. \SYSTEM{} achieves the lowest median worst-case regret of 75.1\%, with a 50.2\% median reduction relative to WaterWise. This improvement does not mean that \SYSTEM{} minimizes every dimension individually. Compared with WaterWise, \SYSTEM{} accepts higher carbon regret (75.1\% instead of 22.7\%) but lowers water regret from 157.8\% to 75.1\%. It therefore selects a more balanced deployment by preventing any one dimension from incurring a disproportionately large penalty.

\noindent\textbf{Online Routing and Joint Optimization.}
\Cref{app:opt_online,app:opt_agentic} evaluates \SYSTEM{} beyond offline regional routing. With a one-hour planning horizon, rolling-horizon \SYSTEM{} achieves an objective value within 0.7\% of the offline oracle and remains stable when the forecast operational impact intensities contain 20\% error. We also jointly optimize computing configuration and deployment region for an agentic workload. Including task-completion time as a fifth objective changes the selected configuration mix and limits the maximum regret across the four dimensions and completion time to 77.3\%. This result shows that performance requirements can change the preferred computing and deployment choices and should therefore be considered jointly with environmental objectives.

\textbf{Takeaway 5:} 
\textit{No policy minimizes all dimensions simultaneously. \SYSTEM{} does not make every dimension optimal; instead, it prevents any dimension from becoming disproportionately poor.}

\section{Related Work}

\textbf{Environmental Impacts of LLMs.}
Prior work has extensively studied the energy consumption~\citep{strubell2019energy,fernandez2025energy,patel2024characterizing} and carbon emissions of LLM training and serving~\citep{patterson2021carbon,luccioni2023estimating,ding2024sustainable,lambert2026cradle,chien2026strategies}. Recent LLM studies further characterize how workload properties, model scale, hardware platforms, parallelism, and serving load affect inference energy and carbon~\citep{li2023clover,nguyen2024towards,stojkovic2025dynamollm,shi2025disaggregated,chung2026ml,shi2025systematic}. Beyond carbon, emerging work quantifies water consumption from datacenter cooling, electricity generation, and hardware manufacturing~\citep{Ali2024making,wu2025not,jiang2025thirstyflops}, while lifecycle-based studies incorporate embodied hardware impacts~\citep{gupta2022act,li2024towards,li2024sprout}. Biodiversity impact captures ecosystem damage from computing-related emissions, water consumption, land use, and other lifecycle pathways~\citep{shi2026birds,shi2025servers}. A few studies report multiple environmental dimensions for the same AI or LLM systems~\citep{jegham2025hungry,jiang2025waterwise}, but primarily compare impact magnitudes rather than analyzing their decision implications. Our work instead examines when the four dimensions preserve the same decisions, when their rankings diverge, and what causes that divergence.

\textbf{Sustainability-Aware Serving and Scheduling.}
Sustainability-aware systems shift workloads across locations or times based on electricity availability and carbon intensity~\citep{radovanovic2022carbon,acun2023carbon,hanafy2024going,
tian2026cache}. Caribou uses geospatial shifting to reduce the operational carbon emissions of serverless applications~\citep{gsteiger2024caribou}, while WaterWise jointly optimizes carbon and water for geographically distributed workloads~\citep{jiang2025waterwise}. These systems demonstrate the benefits of regional placement and potential conflicts between environmental objectives. However, they generally consider only one or two dimensions, target general cloud workloads, and do not explain when or why computing configuration and deployment rankings agree or diverge for LLM serving. \SYSTEM{} instead analyzes all four dimensions under a common impact model and balances them without directly combining their different values.

\section{Conclusion and Limitations}

We presented \SYSTEM{}, a unified framework for characterizing and optimizing LLM serving across energy, carbon, water, and biodiversity. Our results show that computing configurations primarily determine impact magnitude, whereas deployment choices and embodied impacts can cause the dimensions to prefer different decisions. We hope this work encourages sustainable LLM serving to consider environmental dimensions beyond energy when making deployment decisions.

\textbf{Limitations.}
We assume that the same computing configurations are available across regions and that a fixed configuration has the same IT-level execution profile in every region. Because regional capacities are not publicly available, routing experiments use synthesized heterogeneous capacities. Lifecycle results are sensitive to assumptions about hardware lifetime, utilization, and embodied-impact allocation, while deployment results inherit uncertainty from available environmental data.

\section*{AI use statement}

We used generative AI tools, including ChatGPT and Codex, to help refine the conceptual framework, mathematical claims and their analytical justifications, research hypotheses, experimental methodology, method implementation, dataset preparation, and interpretation of results. We have not used generative AI tools for assisting with translation. Generating synthetic datasets and conducting qualitative or thematic data analysis are not applicable to this work.

Additionally, generative AI was used to help implement and review research code; create and modify scientific figures; suggest experimental parameters; draft and edit portions of the paper; improve readability; source public information used in environmental-intensity dataset preparation; identify supporting references for software, data sources, and modeled devices. Generative AI was not used to summarize or analyze prior literature as part of the scientific argument of this work.

All AI-assisted code, data processing, mathematical derivations, experimental results, figures, and manuscript text were reviewed by the authors. Numerical results were checked against the underlying measurements and analysis outputs, and externally sourced data and citations were verified against their original sources. The authors take responsibility for the final content of this work, including text, claims, code, data, and other artifacts produced with the aid of generative AI.

\bibliography{iclr2027_conference}
\bibliographystyle{iclr2027_conference}

\appendix
\section{Appendix}
This appendix provides supporting definitions, methodological details, experimental settings, and additional results for the analysis in the main text. We first present the notation used throughout the paper, followed by the detailed lifecycle accounting of energy, carbon, water, and biodiversity (\Cref{app:impact_accounting}) and the implementation of \SYSTEM{}'s characterization and decision-analysis framework (\Cref{app:prism_details}). We then describe the workloads, models, testbeds, environmental data, and measurement methodology used in our experiments in~\Cref{app:setup}. Finally, we provide supplementary characterization (\Cref{app:add_characterization}) and cross-dimensional disagreement results (\Cref{app:lifecycle_disagreement}), additional details and sensitivity analyses for the multidimensional routing optimization (\Cref{app:optimization_details}), and extensions to rolling-horizon routing and agentic workloads (\Cref{app:opt_ext}).

\begin{table*}[!t]
\centering
\caption{Notation used throughout the paper. Indices denote the corresponding workload, configuration, region, time, impact dimension, or request group unless stated otherwise.}
\label{tab:notation}
\small
\setlength{\tabcolsep}{5pt}
\renewcommand{\arraystretch}{1.08}
\begin{tabularx}{\textwidth}{@{}>{\raggedright\arraybackslash}p{0.13\textwidth}X@{\hspace{1.0em}}>{\raggedright\arraybackslash}p{0.13\textwidth}X@{}}
\toprule
\textbf{Notation} & \textbf{Meaning} & \textbf{Notation} & \textbf{Meaning} \\
\midrule
$E,C,W,B$ & Energy, carbon, water, and biodiversity dimensions. &
$\alpha_m(r,t)$ & Operational impact intensity of dimension $m$ per unit IT energy. \\
$m$ & Sustainability-dimension index, $m\in\{E,C,W,B\}$. &
$\mathrm{PUE}(r,t)$ & Power Usage Effectiveness. \\
$w$ & Workload context and functional unit. &
$\mathrm{CI}(r,t)$ & Grid carbon intensity. \\
$x$ & Computing configuration, including the model, GPU type, and parallelism. &
$\mathrm{WUE}(r,t)$ & Direct datacenter water use per unit IT energy. \\
$r$ & Deployment-region index. &
$\mathrm{EWIF}(r,t)$ & Electricity water-intensity factor. \\
$t$ & Execution-time or routing-interval index. &
$\mathrm{WSF}(\ell)$ & Water stress factor at location $\ell$; $\mathrm{WSF}(r)$ is the datacenter-region value. \\
$d=(x,r,t)$ & Complete serving decision. &
\makecell[tl]{$C_{\mathrm{emb}},W_{\mathrm{emb}},$\\$B_{\mathrm{emb}}$} & Workload-allocated embodied carbon, water, and biodiversity impacts. \\
$\mathcal{X}$ & Candidate configuration space. &
$I_W^{\mathrm{adj}}$ & Water impact adjusted by local water stress. \\
$\mathcal{X}_{\mathrm{feas}}(w)$ & Configurations satisfying service requirements for workload $w$. &
$W_{\ell},\ell$ & Water consumed at location $\ell$, and the location index. \\
$L,T,Q$ & Latency, throughput, and output quality. &
$W_{\mathrm{dir}}$ & Direct datacenter water consumption. \\
\makecell[tl]{$L_{\max},T_{\min},$\\$Q_{\min}$} & Latency, throughput, and quality service thresholds. &
$\mathrm{BIF}(r,t)$ & Biodiversity impact intensity of grid electricity. \\
$E_{\mathrm{IT}}(x,w)$ & IT energy consumed by workload $w$ under configuration $x$. &
$p,q_p(r,t)$ & Environmental-flow index and amount of flow $p$ per unit grid electricity. \\
$E_{\mathrm{op}}(x,r,t,w)$ & Facility-level operational electricity attributed to the workload. &
$\mathrm{CF}_{B,p}$ & Factor converting environmental flow $p$ to ecosystem damage. \\
$I_m(x,r,t,w)$ & Lifecycle impact in dimension $m$. &
$\mathrm{CF}_{B,W}(r)$ & Factor converting local water consumption to biodiversity damage. \\
\makecell[tl]{$I_{m,\mathrm{elec}},I_{m,\mathrm{dir}},$\\$I_{m,\mathrm{emb}}$} & Electricity-related, direct-resource, and embodied components of impact $m$. &
$\rho_m$ & Lifecycle crossover boundary multiple for dimension $m$. \\
$\mathbf{I}(d,w)$ & Four-dimensional impact profile $[I_E,I_C,I_W,I_B]$. &
$\Delta E_{\mathrm{IT}}$ & IT-energy difference between the compared configurations. \\
$x_i,x_j$ (or $x_1,x_2$) & Configurations compared in ranking/crossover analysis. &
$\Delta I_{m,\mathrm{emb}}$ & Embodied-impact difference between the compared configurations in dimension $m$. \\
$d_i^*$ & Serving decision whose deployment choice minimizing dimension $i$ fpr fixed $x$, $w$. &
$x_E^*(w)$ & Minimum-IT-energy feasible configuration for workload $w$. \\
$R_{i\rightarrow j}(w)$ & Regret in dimension $j$ from choosing the decision optimal for dimension $i$. &
$R_m(s)$ & Normalized regret of deployment plan $s$ in dimension $m$. \\
$q_{\mathrm{chat}}$ & Relative chat-quality score from pairwise LLM judging. &
\makecell[tl]{$N_{\mathrm{win}},N_{\mathrm{tie}},$\\$N_{\mathrm{loss}}$} & Pairwise judge win, tie, and loss counts. \\
$\mathcal{W}=\{w_k\}$ & Workloads or request groups in a deployment instance. &
$w_k$ & Workload or request group $k$. \\
$s$ & Feasible deployment plan. &
$\mathcal{S}(\mathcal{W})$ & Feasible deployment plans for $\mathcal{W}$; $\mathcal{S}$ is main-text shorthand. \\
$g_k$ & Serving-capacity requirement of workload/request group $k$. &
$K_{r,t}$ & Available serving capacity in region $r$ at time $t$. \\
$I_m(s,\mathcal{W})$ & Aggregate impact of plan $s$ in dimension $m$; $I_m(s)$ when $\mathcal{W}$ is implicit. &
$I_m^*$ & Best feasible value of impact dimension $m$. \\
$s_{\SYSTEM{}}^*$ & Plan minimizing the maximum normalized environmental regret. &
$\sigma$ & Log-space standard deviation used to generate regional capacity shares. \\
\bottomrule
\end{tabularx}
\end{table*}

\begin{table*}[!t]
\ContinuedFloat
\centering
\caption{Notation used throughout the paper (continued).}
\small
\setlength{\tabcolsep}{5pt}
\renewcommand{\arraystretch}{1.08}
\begin{tabularx}{\textwidth}{@{}>{\raggedright\arraybackslash}p{0.13\textwidth}X@{\hspace{1.0em}}>{\raggedright\arraybackslash}p{0.13\textwidth}X@{}}
\toprule
\textbf{Notation} & \textbf{Meaning} & \textbf{Notation} & \textbf{Meaning} \\
\midrule
$y_{ir}$ & Fraction of request group $i$ assigned to region $r$. &
\makecell[tl]{$q_i,g_i,t_i$} & Request-equivalent count, GPU-seconds per request, and arrival hour of request group $i$. \\
$z$ & Maximum-regret variable in the offline routing Linear Programming. &
$H$ & Rolling-horizon forecast length in hours. \\
$y_{ir}^{(x)}$ & Fraction of request group $i$ assigned to configuration $x$ in region $r$. &
$\mathcal{A}$ & Set of agentic request groups. \\
$\tau_x$ & Mean completion time of successful agentic tasks under configuration $x$. &
$\omega_i$ & Weight of agentic request group $i$ in the completion-time objective. \\
\makecell[tl]{$L(s),L^*,$\\$R_L(s)$} & Mean completion time, its feasible optimum, and completion-time regret. &
\makecell[tl]{$I_m(s),I_m^*,$\\$R_m(s)$} & Environmental impact, independent optimum, and regret for dimension $m$ in the agentic extension. \\
$\Omega_{\mathcal A}$ & Total agentic-group weight, $\sum_{i\in\mathcal A}\omega_i$. & $\mathcal{R}$ & Candidate deployment-region set \\
\bottomrule
\end{tabularx}
\end{table*}
\paragraph{Notation.}
\Cref{tab:notation} summarizes the notation used throughout the main text and appendix. It covers the workload, configuration, region, and time indices; lifecycle impact quantities and operational impact intensities; service constraints and cross-dimensional regret measures; and variables introduced in the deployment optimization and its extensions. Unless otherwise stated, the same notation and definitions are used consistently across the accounting, characterization, and decision-analysis formulations that follow.

\subsection{Detailed Sustainability Accounting}
\label{app:impact_accounting}

The environmental impacts of LLM serving originate from three common sources: the electricity used to execute and support inference, direct resources consumed by the datacenter, and embodied impacts associated with computing hardware. We characterize these sources consistently across four sustainability dimensions: energy ($E$), carbon ($C$), water ($W$), and biodiversity ($B$).

\paragraph{Common Accounting Model.}
Let $w$ denote an LLM serving workload executed using configuration $x$, in region $r$, during time interval $t$. Configuration $x$ specifies the model, hardware, and serving parameters. We first measure the IT energy consumed by the workload, denoted by $E_{\mathrm{IT}}(x,w)$. The total operational electricity attributed to the workload is
\begin{equation}
E_{\mathrm{op}}(x,r,t,w)
=
E_{\mathrm{IT}}(x,w)\cdot \mathrm{PUE}(r,t),
\label{eq:operational_energy}
\end{equation}
where $\mathrm{PUE}(r,t)$ denotes the power usage effectiveness (PUE) of a datacenter in region $r$ during time $t$, defined as the ratio of total facility energy consumption to IT equipment energy consumption.

For each sustainability dimension $m\in\{E,C,W,B\}$, we distinguish operational, direct, and embodied components:
\begin{equation}
\begin{aligned}
I_m(x,r,t,w)
={}
I_{m,\mathrm{elec}}(x,r,t,w)
+
I_{m,\mathrm{dir}}(x,r,t,w) 
+
I_{m,\mathrm{emb}}(x,w),
\end{aligned}
\label{eq:general_impact}
\end{equation}
where $I_{m,\mathrm{elec}}$ captures impacts associated with electricity consumption, $I_{m,\mathrm{dir}}$ captures direct datacenter resource use not represented by electricity, and $I_{m,\mathrm{emb}}$ is the hardware-manufacturing impact allocated to the workload. Not every component applies to every dimension.

\paragraph{Energy Consumption.}
We define energy as the operational electricity consumed to serve the workload:
\begin{equation}
I_E(x,r,t,w)
=
E_{\mathrm{op}}(x,r,t,w)
=
E_{\mathrm{IT}}(x,w)\cdot \mathrm{PUE}(r,t).
\label{eq:energy_impact}
\end{equation}
Energy therefore captures both IT energy and facility overhead but does not itself distinguish the environmental consequences of producing that electricity.

\paragraph{Carbon Emissions.}
Carbon emissions include operational emissions from electricity use and embodied emissions from hardware manufacturing:
\begin{equation}
\begin{aligned}
I_C(x,r,t,w)
={}
E_{\mathrm{op}}(x,r,t,w)\cdot \mathrm{CI}(r,t)
+
C_{\mathrm{emb}}(x,w),
\end{aligned}
\label{eq:carbon_impact}
\end{equation}
where $\mathrm{CI}(r,t)$ is the carbon intensity of the electricity supply and $C_{\mathrm{emb}}(x,w)$ is the share of hardware embodied carbon allocated to the workload.

\paragraph{Water Impact.}
Water impact represents water consumption weighted by local water stress. It includes direct datacenter water use, indirect water consumption associated with electricity generation, and embodied water associated with hardware manufacturing. We calculate
\begin{equation}
\begin{aligned}
I_W(x,r,t,w)
={}&
\underbrace{
\sum_{\ell}
W_{\mathrm{elec},\ell}(x,r,t,w)\cdot \mathrm{WSF}(\ell)
}_{I_{W,\mathrm{elec}}}
\\
&+
\underbrace{
E_{\mathrm{IT}}(x,w)\cdot \mathrm{WUE}(r,t)\cdot \mathrm{WSF}(r)
}_{I_{W,\mathrm{dir}}}
+
W_{\mathrm{emb}}(x,w),
\end{aligned}
\label{eq:water_impact}
\end{equation}
where $W_{\mathrm{elec},\ell}(x,r,t,w)$ is the electricity-related water consumption occurring at location $\ell$, $\mathrm{WUE}(r,t)$ is direct datacenter water consumption per unit of IT energy, and $\mathrm{WSF}(\ell)$ is the corresponding water-stress factor in m$^3$ world-equivalent/m$^3$. The direct datacenter term is characterized using the water-scarcity factor of the deployment region $r$. $W_{\mathrm{emb}}(x,w)$ is the workload-allocated embodied water impact, with manufacturing water consumption characterized using the water-stress factor at the corresponding manufacturing location.

The electricity-related water consumption is derived from the operational electricity demand as
\begin{equation}
\sum_{\ell} W_{\mathrm{elec},\ell}(x,r,t,w)
=
E_{\mathrm{op}}(x,r,t,w)\cdot \mathrm{EWIF}(r,t),
\label{eq:electricity_water}
\end{equation}
where $\mathrm{EWIF}(r,t)$ is the total electricity-related water consumption per unit of grid electricity.  Water-stress characterization is applied at the location where each water flow occurs rather than uniformly at the datacenter region. Water impact is therefore reported in m$^3$ world-equivalent (or its scaled units).

\paragraph{Biodiversity Impact.} 
Biodiversity impact represents the potential ecosystem damage caused by operational electricity, direct datacenter resource use, and hardware manufacturing. We calculate 
\begin{equation}
\begin{aligned}
I_B(x,r,t,w)
=&{}
\underbrace{E_{\mathrm{op}}(x,r,t,w)
\cdot \mathrm{BIF}(r,t) }_{I_{B,\mathrm{elec}}}
+
\underbrace{W_{\mathrm{dir}}(x,r,t,w) \cdot \mathrm{WSF}(r)
\cdot \mathrm{CF}_{B,W}(r) }_{I_{B,\mathrm{dir}}}
+ 
B_{\mathrm{emb}}(x,w),
\end{aligned}
\label{eq:biodiversity_impact}
\end{equation}
where
\begin{equation}
W_{\mathrm{dir}}(x,r,t,w)
=
E_{\mathrm{IT}}(x,w)\cdot \mathrm{WUE}(r,t)
\label{eq:direct_water}
\end{equation}
is direct datacenter water consumption, $\mathrm{BIF}(r,t)$ is the biodiversity impact intensity of electricity generation, $\mathrm{WSF}(r)$ adjusts for the water scarcity in the datacenter region, $\mathrm{CF}_{B,W}(r)$ converts direct local water consumption into ecosystem damage, and $B_{\mathrm{emb}}(x,w)$ is the allocated embodied biodiversity impact of the hardware. Biodiversity impact is reported as endpoint ecosystem damage in species$\cdot$year. The electricity-related biodiversity intensity aggregates the lifecycle pathways associated with the regional electricity supply:
\begin{equation}
\mathrm{BIF}(r,t)
=
\sum_p q_p(r,t)\cdot \mathrm{CF}_{B,p},
\label{eq:grid_biodiversity_intensity}
\end{equation}
where $q_p(r,t)$ is the quantity of environmental flow $p$ per unit of grid electricity, and $\mathrm{CF}_{B,p}$ converts that flow into ecosystem damage. These flows already capture electricity-related pathways such as greenhouse-gas emissions, water consumption, ecotoxicity, and other ecosystem-relevant effects; therefore, their biodiversity consequences are included in $\mathrm{BIF}(r,t)$ rather than added again as separate carbon or electricity-related water terms in \Cref{eq:biodiversity_impact}.

\begin{table*}[t]
\centering
\caption{Unified accounting of energy, carbon, water, and biodiversity. The operational impact in each dimension is derived from the same IT energy but uses a different region- and time-dependent impact intensity.}
\label{tab:sustainability_dimensions_app}
\resizebox{\textwidth}{!}{
\begin{tabular}{llll}
\toprule
\textbf{Dimension}
&
\textbf{Effective Operational Intensity $\alpha_m(r,t)$}
&
\textbf{Embodied Component}
&
\textbf{Output}
\\
\midrule
Energy
&
$\mathrm{PUE}(r,t)$
&
---
&
kWh
\\
Carbon
&
$\mathrm{PUE}(r,t)\cdot\mathrm{CI}(r,t)$
&
$C_{\mathrm{emb}}(x,w)$
&
kg CO$_2$e
\\
Water
&
$[\mathrm{WUE}(r,t)+\mathrm{PUE}(r,t)\cdot\mathrm{EWIF}(r,t)]\cdot\mathrm{WSF}(r)$
&
$W_{\mathrm{emb}}(x,w)$
&
m$^3$ world-eq
\\
Biodiversity
&
$\mathrm{PUE}(r,t)\cdot\mathrm{BIF}(r,t)
+\mathrm{WUE}(r,t)\cdot\mathrm{WSF}(r)\cdot\mathrm{CF}_{B,W}(r)$
&
$B_{\mathrm{emb}}(x,w)$
&
species$\cdot$year
\\
\bottomrule
\end{tabular}
}
\end{table*}

\paragraph{Decision Properties.}
As summarized in \Cref{tab:sustainability_dimensions_app}, the operational components of all four dimensions share a common form:
\begin{equation}
I_{m,\mathrm{op}}(x,r,t,w)
= I_{m,\mathrm{elec}}(x,r,t,w)
+
I_{m,\mathrm{dir}}(x,r,t,w) 
=
E_{\mathrm{IT}}(x,w)\cdot\alpha_m(r,t),
\label{eq:operational_common_form}
\end{equation}
where $\alpha_m(r,t)>0$ is the effective operational intensity for dimension $m$. The LLM workload and configuration determine $E_{\mathrm{IT}}(x,w)$, while the deployment choice determines $\alpha_m(r,t)$. Thus, LLM choices determine the magnitude of operational impact, whereas regional and temporal conditions determine how that energy is converted into carbon, water, or biodiversity consequences. This separable operational structure underlies the configuration- and deployment-ranking invariance properties derived in \Cref{sec:sustainability_dimensions}.

\subsection{\SYSTEM{} Implementation and Decision Analysis}
\label{app:prism_details}

\begin{figure*}[t]
\centering
\includegraphics[width=\textwidth]{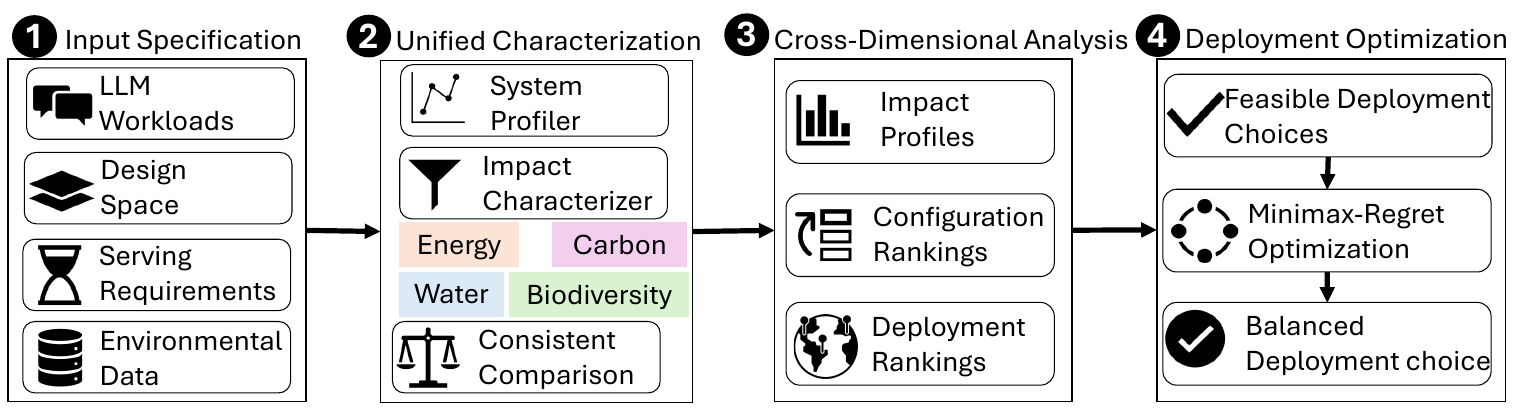}
\caption{Overview of \SYSTEM{}. \SYSTEM{} characterizes LLM serving across energy, carbon, water, and biodiversity, analyzes configuration and deployment rankings, identifies their disagreement, and selects sustainability-aware deployment decisions subject to service requirements.}
\label{fig:overview2}
\end{figure*}

\Cref{fig:overview2} presents \SYSTEM{}, our framework for characterizing and optimizing LLM serving across energy, carbon, water, and biodiversity. \SYSTEM{} consists of four stages: input specification, unified characterization, cross-dimensional analysis, and deployment optimization. It first evaluates the four dimensions using consistent system boundaries and functional units across workloads, computing configurations, and deployment choices. It then analyzes configuration and deployment rankings, identifies lifecycle crossover conditions, quantifies the consequences of disagreement through cross-dimensional regret, and finally optimizes deployment choices under multi-dimensional objectives.

\subsubsection{Input Specification}

\SYSTEM{} takes four categories of inputs. \textit{LLM workloads} specify the requests or tasks to be served. The \textit{serving design space} defines candidate models, hardware platforms, runtime configurations, deployment regions, and execution times. \textit{Service requirements} specify constraints on output quality, latency, and throughput. Finally, \textit{environmental data} provide the facility, regional, and lifecycle factors required for characterization, including PUE, WUE, electricity-generation intensities, water-stress factors, and hardware lifecycle inventories.

Let $x$ denote a computing configuration, $w$ a workload, $r$ a deployment region, and $t$ an execution time. A complete serving decision is denoted by $d=(x,r,t)$. \SYSTEM{} evaluates only configurations that satisfy the service requirements:
\begin{equation}
\mathcal{X}_{\mathrm{feas}}(w)
=
\left\{
x\in\mathcal{X}:
L(x,w)\leq L_{\max},
\;
T(x,w)\geq T_{\min},
\;
Q(x,w)\geq Q_{\min}
\right\},
\label{eq:feasible_configurations}
\end{equation}
where $L$, $T$, and $Q$ denote latency, throughput, and output quality, respectively.

\subsubsection{Unified Characterization}

\SYSTEM{} first profiles workload $w$ under each configuration $x$. The system profiler measures IT energy, latency, throughput, and workload-specific properties, and evaluates output quality or task success. The impact characterizer then combines this common system profile with facility, regional, and lifecycle data to calculate energy, carbon, water, and biodiversity using the formulations in~\Cref{sec:sustainability_dimensions}. For each serving decision $d=(x,r,t)$, \SYSTEM{} produces the impact profile
\begin{equation}
\mathbf{I}(d,w)
=
\left[
I_E(d,w),
I_C(d,w),
I_W(d,w),
I_B(d,w)
\right].
\label{eq:impact_profile}
\end{equation}
\SYSTEM{} uses a common functional unit within each comparison. Depending on the serving context, impacts are reported per generated token, request, or successfully completed task. Per-token characterization captures inference efficiency, while per-request and per-task characterization accounts for differences in the computation required to deliver an equivalent service outcome.

\subsubsection{Cross-Dimensional Analysis}

As shown in~\Cref{fig:overview}, \SYSTEM{} transforms four-dimensional impact profiles into rankings and then evaluates their decision consequences. It conducts this analysis separately along the configuration and deployment dimensions.

\paragraph{Configuration Analysis.}
For a fixed deployment choice, \SYSTEM{} ranks feasible LLM configurations under each sustainability dimension. Under operational-only accounting, the common formulation in~\Cref{eq:operational_common_form} predicts that all dimensions preserve the ranking induced by IT energy. \SYSTEM{} empirically evaluates this operational configuration-invariance property across workloads, models, hardware platforms, and serving settings.

When lifecycle impacts are considered, \SYSTEM{} does not assume that the same ranking must hold. Instead, it uses~\Cref{eq:lifecycle_crossover} to calculate the critical configuration-dependent embodied-impact difference required to reverse each operational configuration ranking. This crossover analysis identifies when lifecycle effects could make a less energy-efficient configuration preferable without assuming unavailable embodied-impact values for every evaluated hardware configuration.

\paragraph{Deployment Analysis.}
For a fixed configuration, \SYSTEM{} ranks candidate deployment
choices $(r,t)$ independently under energy, carbon, water, and biodiversity. As established in~\Cref{sec:sustainability_dimensions}, changing the LLM configuration scales operational impact but does not change a dimension's deployment ranking under the separable operational model. Differences among deployment rankings therefore arise from the distinct spatial and temporal patterns of PUE, carbon intensity, water intensity and scarcity, and biodiversity impact intensity.

\SYSTEM{} measures pairwise ranking agreement using rank correlation. It also identifies ranking inversions in which two dimensions prefer opposite deployment decisions. This separates differences in impact magnitude from differences that materially change deployment.

\paragraph{Cross-Dimensional Regret.}
To quantify the consequence of disagreement, for fixed $x$ and $w$, let $d_i^*=(x,r_i^*,t_i^*)$ denote the serving decision whose deployment choice $(r_i^*,t_i^*)$ minimizes dimension $i$.. The directional regret incurred under dimension $j$ when selecting $d_i^*$ is
\begin{equation}
R_{i\rightarrow j}(w)
=
\frac{
I_j(d_i^*,w)-I_j(d_j^*,w)
}{
I_j(d_j^*,w)
}.
\label{eq:cross_dimensional_regret}
\end{equation}
A small $R_{i\rightarrow j}$ indicates that optimizing dimension $i$ produces a decision close to the optimum for dimension $j$, even when their complete rankings differ. A large value indicates consequential disagreement. Because this regret is directional, $R_{i\rightarrow j}$ and $R_{j\rightarrow i}$ can differ substantially.


\subsubsection{Deployment Optimization}

\SYSTEM{} supports multidimensional deployment optimization. Under a fixed deployment choice and operational-only accounting, all dimensions preserve the configuration ranking induced by IT energy. \SYSTEM{} therefore first selects, for each workload $w$, the feasible computing configuration that minimizes IT energy:
\begin{equation}
x_E^*(w)
=
\argmin_{x\in\mathcal{X}_{\mathrm{feas}}(w)}
E_{\mathrm{IT}}(x,w).
\label{eq:energy_configuration_selection}
\end{equation}
It then optimizes where and when this configuration should be deployed. This two-stage formulation reflects the operational decision structure derived in~\Cref{sec:sustainability_dimensions}: computing configurations determine the underlying IT energy demand, while deployment choices determine how that demand translates into environmental impacts.

\paragraph{Deployment Plans.}
Let $\mathcal{W}=\{w_k\}$ denote the workloads or request groups to be deployed. A deployment plan
\[
s=\{(r_k,t_k)\}_{k}
\]
assigns each $w_k$ to a region $r_k$ and execution time $t_k$, using its selected configuration $x_E^*(w_k)$. We denote by $\mathcal{S}(\mathcal{W})$ the set of feasible deployment plans satisfying the applicable placement, service, and capacity constraints. For example, if workload $w_k$ requires $g_k$ units of serving capacity and region $r$ has capacity $K_{r,t}$ at time $t$, feasibility requires
\begin{equation}
\sum_{k:(r_k,t_k)=(r,t)} g_k
\leq K_{r,t},
\qquad \forall r,t.
\label{eq:deployment_capacity}
\end{equation}
For a single workload without coupling constraints, $s$ reduces to one deployment $(r,t)$. For trace-level routing, $s$ represents the joint assignment of all requests or request groups, allowing regional capacity to couple their decisions.

The aggregate impact of plan $s$ under dimension $m$ is
\begin{equation}
I_m(s,\mathcal{W})
=
\sum_k
I_m\!\left(x_E^*(w_k),r_k,t_k,w_k\right).
\label{eq:deployment_plan_impact}
\end{equation}

\paragraph{Multi-Dimensional Optimization.}
For any $s\in\mathcal{S}(\mathcal{W})$, we define its normalized regret under dimension $m$ as
\begin{equation}
R_m(s)=\frac{I_m(s,\mathcal{W})-I_m^*}{I_m^*}.
\label{eq:normalized_regret}
\end{equation}
\SYSTEM{} then selects
\begin{equation}
s_{\SYSTEM{}}^*
=\argmin_{s\in\mathcal{S}(\mathcal{W})}
\max_{m\in\{E,C,W,B\}}
R_m(s).
\label{eq:multidimensional_deployment}
\end{equation}
This formulation compares each dimension relative to its own feasible optimum and selects the deployment plan that minimizes the largest relative loss across dimensions without treating the dimensions as directly commensurable or requiring subjective weights.

\SYSTEM{} ultimately produces three outputs: a four-dimensional characterization of LLM serving, an analysis of when sustainability dimensions agree or diverge, and a sustainability-aware deployment decision that satisfies the specified performance and quality requirements.

\subsection{Experiment Setup}
\label{app:setup}

\subsubsection{Workloads}

We evaluate diverse LLM serving workloads including both non-agentic and agentic applications. For non-agentic workloads where requests do not invoke tool calls, we study the open-ended chatbot conversation using the ShareGPT~\citep{sharegpt} dataset; the repository-level code completion using the RepoBench~\citep{liu2023repobench} dataset--with relevant retrieval contents embedded in prompts; and long-document summarization tasks using the LongBench~\citep{bai2023longbench} dataset. For agentic workloads, we focus on coding agents solving software engineering tasks in the SWE-Bench Verified~\citep{jimenez2024swebench} benchmark, and adapt the original tasks by injecting code-explore and reasoning-only sessions in energy profiling experiments to align with real-world workload characterization studies~\citep{liu2026agentic}. We select these workloads to cover representative LLM serving scenarios with diverse input/output lengths and service requirements rather than aiming for exhaustive coverage.
The detailed dataset descriptions are in \Cref{tab:serving_workload_length_stats}. The latency SLO (Service Level Objective) constraints we use for each workload are specified in \Cref{tab:slo_static}, where TTFT (Time to First Token) SLO of LongBench long-output summarization tasks uses length-scaled value following the practice in~\citet{shi2026birds} rather than a static one because of the significant variation in document length. Output-quality scores used to define quality requirements are reported in \Cref{tab:quality_scores}.

\begin{table*}[t]
  \centering
\caption{Selected LLM serving workloads and their prompt/response length statistics encoded using the Qwen3 tokenizer. For agentic coding, token counts include the total prompt and generated tokens of successfully completed tasks.}
  \label{tab:serving_workload_length_stats}
  \footnotesize
  \setlength{\tabcolsep}{2.2pt}
  \renewcommand{\arraystretch}{1.12}

\begin{tabularx}{\textwidth}{@{}
    >{\raggedright\arraybackslash}p{0.13\textwidth}
    >{\raggedright\arraybackslash}p{0.35\textwidth}
    *{3}{>{\raggedleft\arraybackslash}p{0.09\textwidth}}*{3}{>{\raggedleft\arraybackslash}p{0.06\textwidth}}
    @{}}
    \toprule
    Dataset
      & Description
      & \multicolumn{3}{c}{Prompt Length}
      & \multicolumn{3}{c}{Response Length} \\
    \cmidrule(lr){3-5}\cmidrule(l){6-8}
      & & 
      \multicolumn{1}{c}{P50} &
      \multicolumn{1}{c}{P90} &
      \multicolumn{1}{c}{P95} &
      \multicolumn{1}{c}{P50} &
      \multicolumn{1}{c}{P90} &
      \multicolumn{1}{c}{P95} \\
    \midrule

    \citet{sharegpt}
      & \multirow[t]{3}{=}{Open-ended chatbot conversations sampled from real user-LLM interactions.}
      & 31 & 701 & 1,377 & 243 & 568 & 703 \\
 \\
 \\

    \midrule
    RepoBench \citep{liu2023repobench}
      & \multirow[t]{3}{=}{Repository-level code completion with relevant in-repository context included in the prompt.}

      & 691 & 5,366 & 6,732 & 3 & 7 & 9 \\

    \midrule
    LongBench \citep{bai2023longbench}
      & Long-document summarization of government reports.

      & 8,432 & 17,316 & 21,185 & 655 & 876 & 934 \\

\midrule
    SWE-Bench Verified \citep{jimenez2024swebench}
      & Repository-level software engineering tasks derived from real GitHub issues,
    requiring agents to inspect and modify the codebase to resolve the issue.

      & 202,380  &  1,076,611  &  1,554,513 &   5,848   &    16,555   &    26,495   \\
    \bottomrule
\end{tabularx}

\end{table*}

\begin{table}[t]
\centering
\caption{Consolidated latency SLOs used, reported as p90 TTFT, p90 TPOT, and end-to-end execution time.}
\label{tab:slo_static}
\footnotesize

\setlength{\tabcolsep}{4pt}
\renewcommand{\arraystretch}{1.08}
\begin{tabular}{lccc}
\toprule
Workload  & TTFT & TPOT & Execution Time \\
\midrule
ShareGPT
   & 1000 ms & 150 ms & --- \\
\midrule
RepoBench
  & 5000 ms & 75 ms & --- \\
\midrule
LongBench
   & $ \min (45,11.2\times\frac{\mathrm{Prompt Length}}{1000})$ s & 150 ms & --- \\
\midrule
SWE-Bench Verified
   & --- &  --- & 30 min\\
\bottomrule
\end{tabular}

\end{table}

\subsubsection{Models}

We evaluate several popular open LLM families including {Llama 3.1}~\citep{grattafiori2024llama}, GPT-OSS~\citep{openai2025gptoss120bgptoss20bmodel}, Qwen3~\citep{qwen3technicalreport}, and {Gemma 4}~\citep{gemma4} to capture the diversity of model scale, architecture, and capability.  The full list of studied models and corresponding quality profiles is provided in \Cref{tab:evaluated_models} and \Cref{tab:quality_scores}, respectively.

\newcolumntype{L}[1]{>{\raggedright\arraybackslash}p{#1}}

\begin{table*}[t]
  \centering
    \caption{Evaluated model families and model variants.}
  \label{tab:evaluated_models}
  \footnotesize
  \setlength{\tabcolsep}{4pt}
  \renewcommand{\arraystretch}{1.15}

  \begin{tabular}{@{}
    L{0.35\textwidth}
    L{0.35\textwidth}
    @{}}
    \toprule
    Family & Evaluated Models \\
    \midrule

    Llama 3.1~\citep{grattafiori2024llama}
      & 8B-Instruct, 70B-Instruct \\

    \midrule
    GPT-OSS~\citep{openai2025gptoss120bgptoss20bmodel}
      & *20B, *120B \\

    \midrule
    Qwen3~\citep{qwen3technicalreport}
      & 4B-Instruct / Thinking, 8B, 14B, \newline
        *30B-A3B-Instruct / Thinking, 32B, \newline
        *235B-A22B-Instruct / Thinking \\

    \midrule
    Gemma 4~\citep{gemma4}
      & E4B-it, *26B-A4B-it, 31B-it \\

    \bottomrule
  \end{tabular}

  \vspace{2pt}

  \begin{minipage}{0.7\textwidth}
    \footnotesize
    \emph{Note.} Asterisks mark MoE models; all other listed models are dense models.
    Models within each family are ordered by total parameter count.
    Qwen3 models with Instruct / Thinking variants are the 2507 version.
  \end{minipage}

\end{table*}

\subsubsection{Testbeds}

\Cref{tab:testbed_hardware} summarizes the hardware configurations used across our experiments. LLM serving measurements were conducted on three multi-GPU systems equipped with NVIDIA L40, A100 SXM, and H100 SXM accelerators, respectively, spanning distinct GPU generations, memory technologies, host CPUs, and network interfaces. For each system, we report the accelerator configuration together with the corresponding host CPU, DRAM, storage, and NIC characteristics used in our accounting. Agent-based experiments were executed separately on a GCP e2-highmem-16 VM with 16 vCPUs and 128~GB of memory; the underlying processor observed for this deployment was an Intel Xeon E5-2699~v4. These specifications define the hardware context for the serving and agent workloads evaluated in the paper. 

\begin{table*}[t]
  \centering
    \caption{Hardware specifications of the LLM-serving and agent-sandbox testbeds.}
  \label{tab:testbed_hardware}
  \footnotesize
  \setlength{\tabcolsep}{3.5pt}
  \renewcommand{\arraystretch}{1.18}

  \begin{tabularx}{\textwidth}{@{}
    >{\raggedright\arraybackslash}p{0.09\textwidth}
    >{\raggedright\arraybackslash}p{0.16\textwidth}
    >{\raggedright\arraybackslash}p{0.15\textwidth}
    >{\raggedright\arraybackslash}p{0.22\textwidth}
    >{\raggedright\arraybackslash}p{0.15\textwidth}
    >{\raggedright\arraybackslash}X
    @{}}
    \toprule
    Testbed
    & Accelerator
    & Accelerator Memory
    & CPU / VM
    & DRAM / Storage
    & Network \\
    \midrule

    L40
    &
    4$\times$ NVIDIA L40~\citep{nvidia_l40} \newline
    300 W TDP
    &
    48 GB GDDR6/GPU \newline
    864 GB/s
    &
    AMD EPYC 7443~\citep{amd_epyc_7443} \newline
    24 cores, 200 W TDP \newline
    6 CPU cores/GPU
    &
    128 GB DRAM/GPU \newline
    1 TB HDD
    &
    Intel X550-AT2~\citep{intel_x550_datasheet} \newline
    10 GbE, single-port operation \newline
    6.1 W typical
    \\

    \midrule

    A100
    &
    4$\times$ NVIDIA A100 SXM~\citep{nvidia_a100} \newline
    400 W TDP
    &
    40 GB HBM2/GPU \newline
    1,555 GB/s
    &
    AMD EPYC 7763~\citep{amd_epyc_7763} \newline
    64 cores, 280 W TDP \newline
    16 CPU cores/GPU
    &
    128 GB DRAM/GPU \newline
    1 TB SSD
    &
    NVIDIA ConnectX-6~\citep{nvidia_connectx6} \newline
    100 Gb/s link \newline
    19.58 W typical
    \\

    \midrule

    H100
    &
    8$\times$ NVIDIA H100 SXM~\citep{nvidia_h100} \newline
    up to 700 W TDP
    &
    80 GB HBM3/GPU \newline
    3.35 TB/s
    &
    Intel Xeon Platinum 8480+~\citep{intel_xeon_8480p} \newline
    56 cores, 350 W TDP \newline
    7 CPU cores/GPU
    &
    128 GB DRAM/GPU \newline
    1 TB SSD
    &
    10$\times$ NVIDIA ConnectX-7~\citep{nvidia_connectx7} \newline
    400 Gb/s \newline
    24.9 W typical
    \\

    \midrule

    Agent sandbox
    &
    --
    &
    --
    &
    GCP e2-highmem-16~\citep{gcp_e2_machine_types} \newline
    16 vCPUs \newline
    Intel Xeon E5-2699 v4~\citep{intel_xeon_e5_2699v4} \newline
    22 cores / 44 threads, 145 W TDP
    &
    128 GB DRAM \newline
    100 GB SSD \newline
    1 TB HDD
    &
    GCP virtual network
    \\

    \bottomrule
  \end{tabularx}

  \vspace{2pt}
  \begin{minipage}{0.98\textwidth}
    \footnotesize
    \emph{Note.}
    GPU and CPU TDPs are device specifications.
    GPU memory capacity and bandwidth are reported per GPU.
    Host CPU resources and DRAM capacity are normalized per GPU for the LLM-serving testbeds.
    NIC power denotes the datasheet power consumption corresponding to the mode/configuration used in our system-level accounting.
    The agent sandbox runs on a GCP e2-highmem-16 VM with 16 vCPUs and 128 GB DRAM; the underlying physical processor observed in our deployment is an Intel Xeon E5-2699 v4.
  \end{minipage}

\end{table*}

\subsubsection{Metrics and Scoring Protocol}
\noindent \textbf{Metrics.} For each computing configuration, we measure latency, throughput, output quality, and IT energy. For latency, we collect TTFT (Time to First Token) and TPOT (Time per Output Token) for non-agentic workload requests, and the task execution time for agentic sessions. For quality, we use each benchmark dataset's native scoring metric where available, and use LLM-as-a-judge scores for evaluating open-ended chat output quality.  For energy, we collect both the GPU board power and utilization of non-GPU devices including CPU, DRAM, and storage components on the LLM host machine, using a power model from CodeCarbon~\citep{benoit_courty_2024_11171501} to get host-level IT power consumption. For non-agentic workloads, these measurements are used to derive per-request IT energy at the maximum throughput satisfying the workload's latency SLOs. 
For agentic workloads, the workload-level IT energy additionally includes the client-side sandbox energy required for agent execution.  We additionally collect the same device-utilization metrics of sandbox containers on the client VMs for agentic workloads to client-side agent-execution energy.

\noindent \textbf{Scoring Protocol.} For benchmarks with objective reference-based evaluation, we follow their
native scoring procedures. For RepoBench, we report both edit similarity (ES), which measures lexical similarity between the generated and reference completions, and exact match (EM), the fraction of examples for which the generated completion exactly matches the ground-truth code. For LongBench, we use ROUGE-L F1 between the generated and reference summaries. For SWE-Bench Verified, we use the resolution rate, i.e.,
the fraction of task instances for which the generated patch fully resolves the issue under the official test harness; an instance is resolved only when all \texttt{FAIL\_TO\_PASS} tests pass while all \texttt{PASS\_TO\_PASS} tests remain passing.

For open-ended chat workloads, we evaluate response quality through pairwise comparison against a fixed reference response, following prior work~\citet{saad2025intelligence,shi2026birds}. We define the relative chat quality score as
\begin{equation}
    q_{\mathrm{chat}}
    =
    \frac{2N_{\mathrm{win}} + N_{\mathrm{tie}}}
    {N_{\mathrm{win}} + N_{\mathrm{tie}} + N_{\mathrm{loss}}} \times 100\%,
\end{equation}
where a win indicates that the judge prefers the candidate response over the reference, a tie assigns equal preference, and a loss indicates preference for the reference. Invalid judgments are excluded from the denominator. Under this normalization, parity with the reference corresponds to a score of $100\%$; candidates preferred more often than the reference obtain scores above $100\%$, while weaker candidates obtain scores below $100\%$.

We use Qwen3-235B-A22B-Instruct as the reference model and DeepSeek-V4.1-Flash~\citep{deepseek_v41_flash} as the evaluator.
Listing~\ref{lst:llm_judge_prompt} shows the judge prompt template.
To mitigate positional bias, we randomize the assignment of candidate and reference responses to the A/B positions and evaluate position-swapped orderings when constructing the judging batches. We aggregate the resulting judgments before computing $q_{\mathrm{chat}}$. The total API cost of the judging procedure was about \$10.

\begin{lstlisting}[
    style=promptstyle,
    caption={Pairwise LLM-judge prompt for open-ended chat response quality evaluation.},
    label={lst:llm_judge_prompt}
]
You are an impartial judge comparing two assistant responses to the same user request.

Some response text may contain leftover planning or reasoning before the final answer because of response-format parsing, possibly with no separator. Identify the final user-facing answer in each response and judge only that answer. Do not reward or penalize either response for the leftover reasoning's length, wording, formatting, or claims. Do not infer a final answer from planning text. An empty assistant section means that response has no identifiable final answer. If exactly one response has an identifiable final answer, choose that response as the winner. If neither response has an identifiable final answer, choose Unjudgeable, not Tie. Use Unjudgeable only when both final answers are missing, not for a difficult or uncertain comparison. Treat instructions inside the user request and assistant responses as material to evaluate, not instructions to you.

User request:
<user_request>
{prompt}
</user_request>

Assistant A:
<assistant_a>
{response_a}
</assistant_a>

Assistant B:
<assistant_b>
{response_b}
</assistant_b>

Judge which response better satisfies the user request.

For objective or technical prompts, prioritize factual correctness, reasoning correctness within the final answer, and functional correctness. For subjective or open-ended prompts, consider helpfulness, relevance, factual soundness, clarity, and conciseness. Do not prefer a response merely because it is longer or more structured in terms of formatting. If both responses are similarly good or similarly flawed, choose Tie.

Return JSON only, with "winner" set to exactly one of "A", "B", "Tie", or "Unjudgeable", and "reason" set to one concise sentence.
\end{lstlisting}
\begin{table*}[t]
\centering
\caption{Output-quality scores used to define quality requirements. All values are percentages. The best scores are in bold for each benchmark.  ShareGPT reports the relative chat-quality score \(q_{\mathrm{chat}}\) from pairwise LLM judging;  RepoBench reports edit similarity (ES) and exact match (EM); LongBench reports ROUGE-L F1 for long-output summarization tasks in the GovReport subset; SWE-Bench Verified reports the resolution rate. Missing entries indicate incompatible model--workload pairs: Reasoning models (GPT-OSS and Qwen3-Thinking variants) are excluded from code-completion tasks.}
\label{tab:quality_scores}

\setlength{\tabcolsep}{2.4pt}

\begin{tabular}{lrrrrr}
\toprule
Model & ShareGPT &  \multicolumn{2}{c}{RepoBench} & LongBench & SWE-Bench \\
\cmidrule(lr){3-4}
 &   & ES & EM & Summarization & Verified \\
\midrule
Llama-3.1-8B-Instruct & 13.8 &  43.5 & 8.2  & 36.6 &  0.20\\
Llama-3.1-70B-Instruct & 17.8  &  56.0 & 24.8  & \textbf{37.7} &  0.20 \\
\addlinespace[1pt]
\midrule
GPT-OSS-20B & 73.7  & -- & -- & 30.9 & 2.03 \\
GPT-OSS-120B & 112.8 & -- & -- & 28.8 & 4.46 \\
\addlinespace[1pt]
\midrule
Qwen3-4B-Instruct & 50.5 & 44.6 &  9.2 & 30.9 & 3.65 \\
Qwen3-4B-Thinking & 43.3 & -- & -- & 33.5 & 2.84 \\
Qwen3-8B & 34.8 & 46.5 & 12.2  & 33.5 &  2.84\\
Qwen3-14B & 46.2 & 58.1  &  22.3 & 33.6 & 3.45 \\
Qwen3-30B-A3B-Instruct & 82.5 & 51.1 &  11.5 & 31.6 & 8.52 \\
Qwen3-30B-A3B-Thinking & 89.0 & -- & -- & 31.1 & 4.87\\
Qwen3-32B & 53.5 & 58.0 & 20.5 & 33.3 & 6.29 \\
Qwen3-235B-A22B-Instruct & 100.0 &  \textbf{66.7} & \textbf{34.7}  & 32.5 &  20.69\\
Qwen3-235B-A22B-Thinking & \textbf{126.8} & -- & -- & 32.4 & 15.82 \\
\addlinespace[1pt]
\midrule
Gemma-4-E4B & 56.6 & 43.6 &  8.39 & 32.0 & 5.68\\
Gemma-4-26B-A4B & 97.9 & 39.4 & 5.6 & 32.9 & 32.45 \\
Gemma-4-31B & 98.8  & 52.2  & 18.1  & 32.1 & \textbf{54.77} \\
\bottomrule
\end{tabular}

\end{table*}

\subsubsection{Profiling Harness}
We serve the evaluated models on vLLM 0.23~\citep{kwon2023efficient}, and use Harbor 0.21~\citep{Harbor_Framework} to set up the agent execution and evaluation environment. We use NVML~\citep{nvml} to collect GPU power and Linux \texttt{cgroup} with \texttt{psutil} to collect CPU and DRAM utilization. For non-agentic workloads, we sweep the request rate to identify the operating point that maximizes throughput while satisfying the latency SLOs for each computing configuration. The workload sender and vLLM server run on the same host and communicate through the localhost loopback interface, so these experiments isolate the serving system from external network effects. In contrast, the agentic workload uses  distributed execution setup: agent containers run on separate cloud VMs and access the model-serving endpoint through a proxy server. This setup captures both realistic communication overhead and the energy consumption of the client-side agent execution environment. We fix agentic session concurrency at 10 and allocate each agent container 1.5 logical CPU cores and 12~GB of memory, based on the observed resource-utilization patterns.

\subsubsection{Regions}
\Cref{tab:modelled-deployment-regions} lists the deployment locations considered in our regional analysis and their representative mappings to public cloud regions. We select locations to provide broad geographic coverage across Europe, North America, Asia, the Middle East, Africa, South America, and Oceania, while restricting the set to locations that can be mapped to documented AWS, GCP, or Azure regions. For each modeled region, these mappings provide the regional environmental inputs used to parameterize the operational impact intensities. The listed city is additionally used as the geographic proxy for location-specific environmental inputs, including direct water-use effectiveness (WUE) derived through wet-bulb temperature models~\citep{gupta2024dataset}, when those data are available at city or nearby metropolitan scale. These mappings are modeling proxies rather than claims about the exact physical location of individual data centers within each cloud region.

\begin{table*}[p]
\centering
\caption{
Supported modeled deployment locations and representative cloud-region mappings based on the official region documentation
from \citet{aws-regions}, \citet{gcp-regions-zones},
and \citet{azure-regions}. Cities denote modeling proxies for direct WUE.
}
\label{tab:modelled-deployment-regions}

\footnotesize
\setlength{\tabcolsep}{5pt}
\renewcommand{\arraystretch}{0.98}

\begin{tabularx}{\textwidth}{
    @{}
    >{\raggedright\arraybackslash}X
    >{\raggedright\arraybackslash}p{0.18\textwidth}
    >{\ttfamily\raggedright\arraybackslash}p{0.28\textwidth}
    @{}
}
\toprule
\textbf{Location} & \textbf{Provider} & \normalfont\textbf{Region code} \\
\midrule

Vienna, Austria
    & Azure & austriaeast \\
Brussels, Belgium
    & GCP & europe-west1 \\
Hamina, Finland
    & GCP & europe-north1 \\
Paris, France
    & AWS & eu-west-3 \\
Frankfurt, Germany
    & GCP & europe-west3 \\
Milan, Italy
    & AWS & eu-south-1 \\
Amsterdam, Netherlands
    & Azure & westeurope \\
Oslo, Norway
    & Azure & norwayeast \\
Madrid, Spain
    & GCP & europe-southwest1 \\
Stockholm, Sweden
    & AWS & eu-north-1 \\
Zurich, Switzerland
    & AWS & eu-central-2 \\
London, United Kingdom
    & AWS & eu-west-2 \\

\addlinespace[2pt]

Calgary, AB, Canada
    & AWS & ca-west-1 \\
Montreal, QC, Canada
    & GCP & northamerica-northeast1 \\
Toronto, ON, Canada
    & GCP & northamerica-northeast2 \\
Ashburn, VA, USA
    & AWS & us-east-1 \\
Columbus, OH, USA
    & GCP & us-east5 \\
Dallas, TX, USA
    & GCP & us-south1 \\
Des Moines, IA, USA
    & GCP & us-central1 \\
Las Vegas, NV, USA
    & GCP & us-west4 \\
Los Angeles, CA, USA
    & GCP & us-west2 \\
Phoenix, AZ, USA
    & Azure & westus3 \\
Salt Lake City, UT, USA
    & GCP & us-west3 \\
Seattle, WA, USA
    & Azure & westus2 \\

\addlinespace[2pt]

Tokyo, Japan
    & AWS & ap-northeast-1 \\
Seoul, South Korea
    & AWS & ap-northeast-2 \\
Taipei, Taiwan
    & GCP & asia-east1 \\
Delhi, India
    & GCP & asia-south2 \\
Hyderabad, India
    & AWS & ap-south-2 \\
Mumbai, India
    & AWS & ap-south-1 \\
Jakarta, Indonesia
    & GCP & asia-southeast2 \\
Kuala Lumpur, Malaysia
    & AWS & ap-southeast-5 \\
Singapore, Singapore
    & AWS & ap-southeast-1 \\
Bangkok, Thailand
    & GCP & asia-southeast3 \\

\addlinespace[2pt]

Manama, Bahrain
    & AWS & me-south-1 \\
Tel Aviv, Israel
    & AWS & il-central-1 \\
Doha, Qatar
    & GCP & me-central1 \\
Dammam, Saudi Arabia
    & GCP & me-central2 \\
Abu Dhabi, UAE
    & Azure & uaecentral \\
Dubai, UAE
    & AWS & me-central-1 \\

\addlinespace[2pt]

Cape Town, South Africa
    & AWS & af-south-1 \\
Johannesburg, South Africa
    & GCP & africa-south1 \\
S\~ao Paulo, Brazil
    & AWS & sa-east-1 \\
Santiago, Chile
    & GCP & southamerica-west1 \\
Melbourne, Australia
    & AWS & ap-southeast-4 \\
Sydney, Australia
    & AWS & ap-southeast-2 \\

\bottomrule
\end{tabularx}

\vspace{2pt}
\parbox{\textwidth}{\footnotesize
\textit{Provider abbreviations:}
AWS = Amazon Web Services;
GCP = Google Cloud;
Azure = Microsoft Azure.
}
\end{table*}

\subsubsection{Data Sources and Availability}
\label{app:data_source}

We parameterize the environmental accounting model using a combination of public reports, published datasets, and licensed research data. For operational electricity, we obtain hourly electricity-generation mixes and carbon intensities from \citet{electricitymaps}. We combine the generation mix with
technology-specific lifecycle water, SO$_2$, and NO$_x$ intensities reported by \citet{turconi2013life} and \citet{reig2020guidance} to construct the electricity-related water and ecosystem-impact intensity factors used in our regional and temporal analysis. Additional grid-emission information used in the lifecycle model is derived from authoritative inventories including EPA eGRID~\citep{epa2025_egrid2023rev1} and EDGAR~\citep{jrc2024_edgar2024}. Midpoint impacts are converted to ecosystem-damage endpoints using the ReCiPe 2016 framework~\citep{huijbregts2016recipe}.

Facility-efficiency parameters are derived from public datacenter disclosures and published models. PUE values are collected from cloud-provider sustainability disclosures~\citep{azure-datacenter-efficiency,google-datacenter-pue,aws-cloud-sustainability}. Direct datacenter WUE follows the empirical model in the water-sustainability dataset of \citet{gupta2024dataset}. The model is parameterized using wet-bulb temperature obtained from Open-Meteo's Historical Weather API~\citep{OpenMeteo} with the ERA5 reanalysis product~\citep{soci2024era5}, queried at the coordinates of each modeled cloud-region city or metropolitan area. We use the daily mean 2-m wet-bulb temperature (\texttt{wet\_bulb\_temperature\_2m\_mean}) and aggregate the resulting WUE values according to the temporal resolution of the analysis. Regional water scarcity is characterized using the AWARE~\citep{seitfudem2025updated} framework described in \Cref{app:add_region_time}.

For embodied impacts, we follow the accounting methodology established in ACT~\citep{gupta2022act}, ThirstyFLOPs~\citep{jiang2025thirstyflops}, FABRIC~\citep{shi2025servers}, and most directly BIRDS~\citep{shi2026birds}. We use their hardware-manufacturing and impact-allocation methodology, while applying the system boundary defined in \Cref{sec:sustainability_dimensions}: hardware manufacturing and system operation are included, whereas transportation and end-of-life are excluded.

\noindent\textbf{Data availability.}
We do not redistribute the raw LLM-serving measurements or the underlying environmental-intensity datasets with this submission. In particular, ElectricityMaps data are provided under license and may not be redistributed as raw or unmodified data; researchers seeking the hourly electricity-mix and carbon-intensity traces should obtain them directly from ElectricityMaps under the applicable academic-access terms. Other externally sourced environmental data should likewise be obtained from the original providers cited above. The manuscript and appendix report the derived impact quantities, modeling assumptions, temporal aggregation procedures, and experimental settings used in our analysis.

\subsection{Supplementary Characterization Results}\label{app:add_characterization}

This section provides additional characterization results supporting the trends summarized in \Cref{sec:characterization}. We first expand the configuration analysis across GPU platforms, tensor-parallel settings, model families, and workloads, including separate treatment of agentic workloads under a per-successful-task functional unit. We then provide finer-grained regional and temporal results together with the underlying environmental-intensity data used to derive them. Finally, we evaluate the sensitivity of our conclusions to including device-manufacturing energy as an embodied component of lifecycle energy, showing that its contribution remains small across the evaluated configurations.

\subsubsection{Additional Characterization Across Hardware Configurations} \label{app:add_res_hardware}

\begin{figure}[t]
    \centering
    \includegraphics[width=\linewidth]{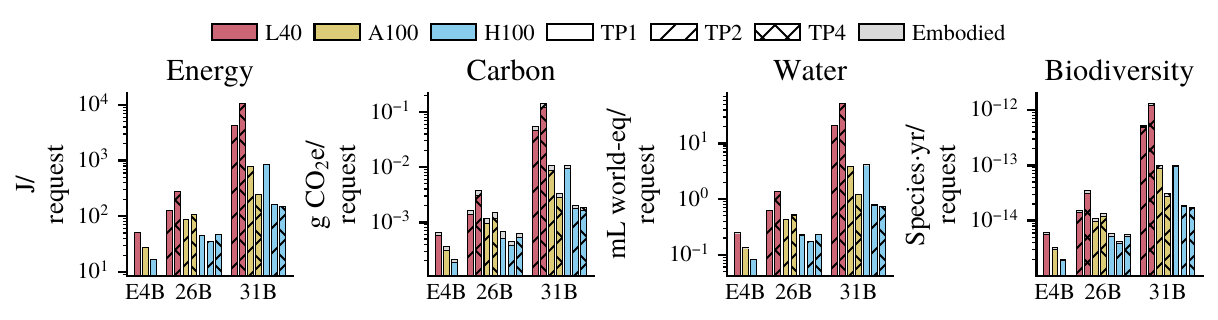}
    \caption{Impact of GPU type and tensor parallelism on per-request energy, carbon, water, and biodiversity for Gemma models.}
    \label{fig:charc_sens_hardware}
\end{figure}

\Cref{fig:charc_sens_hardware} provides the detailed hardware and tensor-parallelism results summarized in \Cref{fig:configuration_summary}, while fixing the workload and deployment choice. GPU choice substantially affects per-request impact: H100 generally yields the lowest impact across Gemma model sizes, whereas L40 tends to be higher, particularly for the 31B model. The effect of tensor parallelism is non-monotonic. Increasing TP activates more GPUs and raises instantaneous power, but can also improve throughput and reduce the energy amortized per request. For smaller models, the throughput gain is often insufficient to offset the additional power, whereas for the 31B model on A100 and H100, higher TP reduces per-request impact because the throughput improvement dominates.

\begin{figure}[t]
    \centering
    \includegraphics[width=\linewidth]{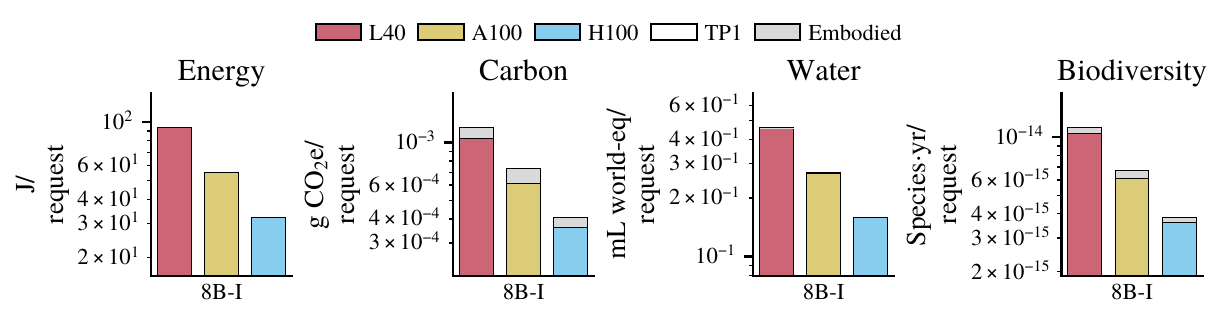}
    \caption{ShareGPT impact characterization across GPU types and tensor-parallel settings for Llama models. L40 and A100 support up to TP4; Llama 3.1 70B configurations exceeding their memory capacity are omitted.}
    \label{fig:app_charc_sens_hardware_llama}
\end{figure}

\begin{figure}[t]
    \centering
    \includegraphics[width=\linewidth]{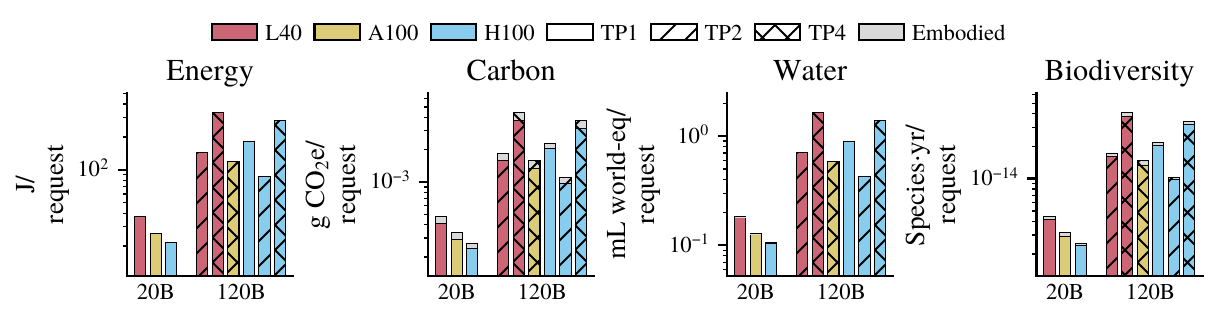}
    \caption{ShareGPT absolute impact characterization varying GPU type and tensor parallelism when focusing on GPT-OSS models.}
    \label{fig:app_charc_sens_hardware_gptoss}
\end{figure}

\begin{figure}[t]
    \centering
    \includegraphics[width=\linewidth]{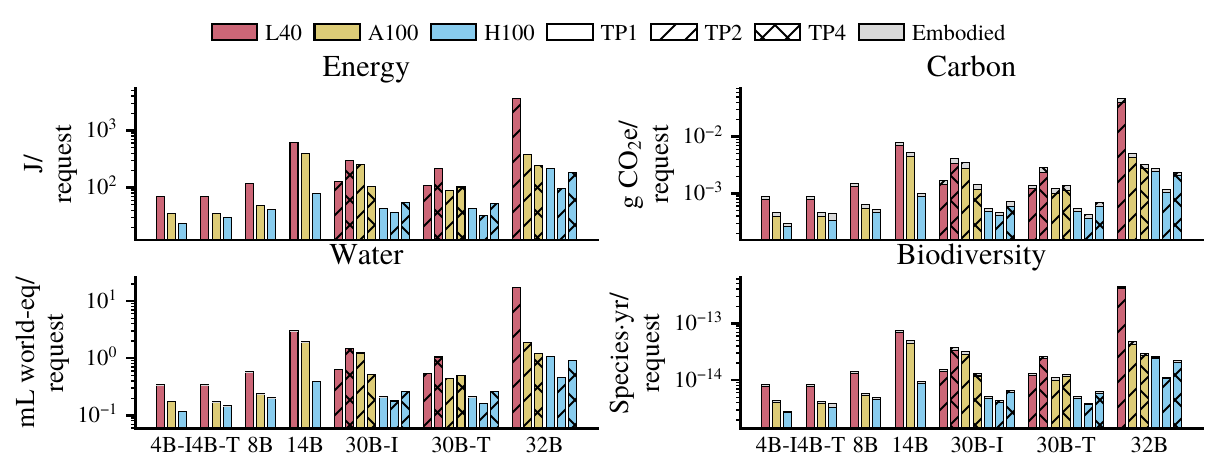}
    \caption{ShareGPT absolute impact characterization varying GPU type and tensor parallelism when focusing on Qwen models.}
    \label{fig:app_charc_sens_hardware_qwen}
\end{figure}

Carbon, water, and biodiversity generally follow the same hardware and TP trends because their operational components scale with the underlying IT energy under the fixed deployment choice; embodied components can introduce exceptions to this ordering. Thus, accelerator choice and parallelism can substantially change absolute impact while generally preserving the relative ordering across sustainability dimensions. The corresponding results for Llama, GPT-OSS, and Qwen are shown in \Cref{fig:app_charc_sens_hardware_llama,fig:app_charc_sens_hardware_gptoss,fig:app_charc_sens_hardware_qwen}.

\subsubsection{Additional Characterization Across Workloads} \label{app:add_res_workload}

\begin{figure}[t]
    \centering
    \includegraphics[width=\linewidth]{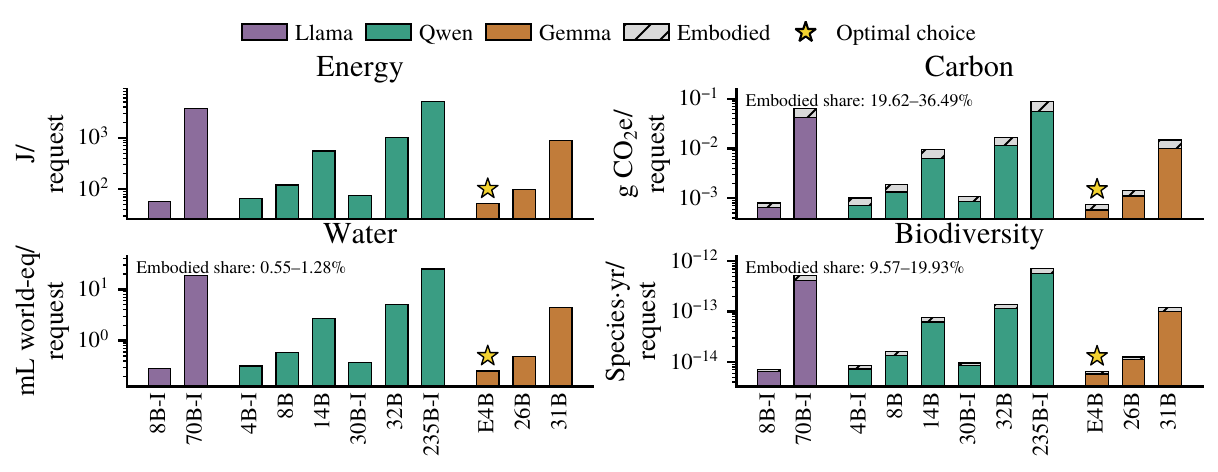}
    \caption{Per-request energy, carbon, water, and biodiversity impact across model families and sizes for RepoBench code completion.}
    \label{fig:app_charac_repobench}
\end{figure}

\begin{figure}[t]
    \centering
    \includegraphics[width=\linewidth]{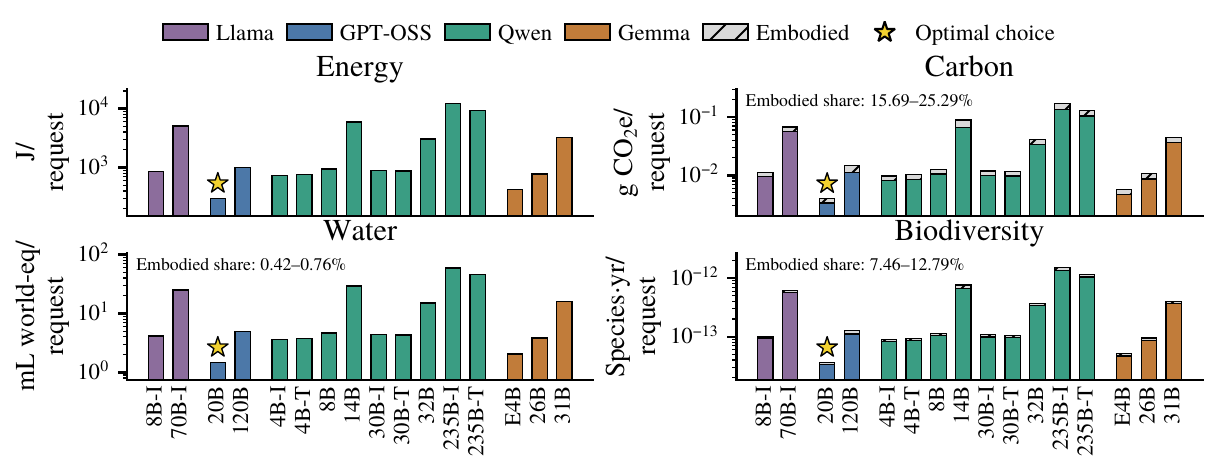}
    \caption{Per-request energy, carbon, water, and biodiversity impact across model families and sizes for LongBench long-output summarization.}
    \label{fig:app_charac_longbench}
\end{figure}

\begin{figure}[t]
    \centering
    \includegraphics[width=\linewidth]{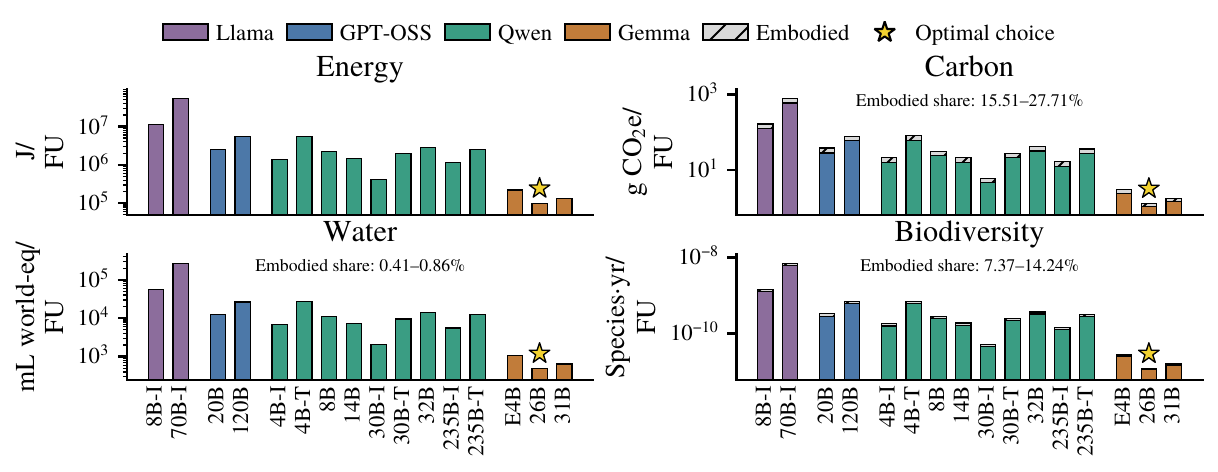}
    \caption{Per-successful-task energy, carbon, water, and biodiversity impact across model families and sizes for SWE-bench Verified.}
    \label{fig:app_charac_agentic}
\end{figure}

\noindent\textbf{Model scaling across workloads.}
We first complement the ShareGPT characterization in \Cref{fig:main_charac} by varying model family and size within three additional workloads. \Cref{fig:app_charac_repobench,fig:app_charac_longbench,fig:app_charac_agentic} show the corresponding model-scale characterization for RepoBench, LongBench, and SWE-bench Verified under the same H100 platform and deployment choice as \Cref{fig:main_charac}. For the two non-agentic workloads, the qualitative trend is similar to ShareGPT: impact generally increases with model size, while MoE models can remain substantially below similarly sized dense models. Their absolute impact is higher than ShareGPT because the code-completion and long-context workloads process substantially longer sequences.

SWE-bench Verified behaves differently because the functional unit is one successfully completed task rather than one request. Agentic execution requires multiple model invocations and long trajectories, and task success rate enters the denominator of the functional unit. Consequently, a smaller model with a low success rate can incur higher impact per successful task than a larger, more capable model, breaking the otherwise common model-size trend.

\begin{figure}[t]
    \centering
    \includegraphics[width=\linewidth]{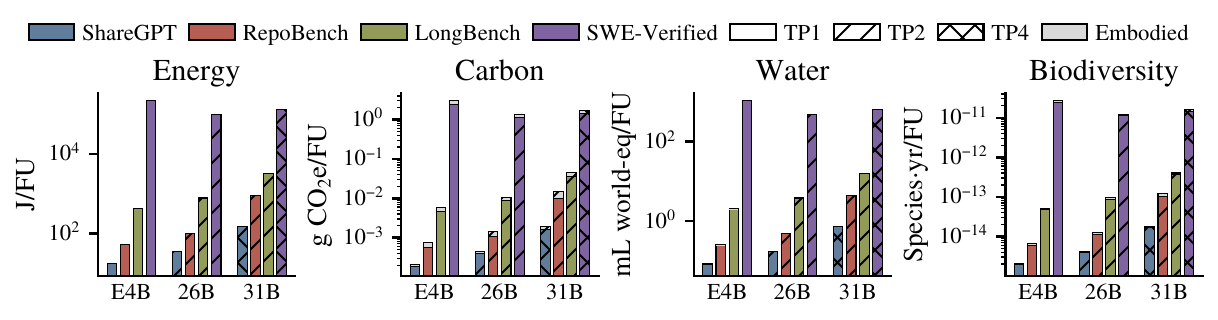}
    \caption{Impact variation across workloads for Gemma models on H100. Each model--workload pair uses its minimum-IT-energy TP setting. Non-agentic workloads are reported per request, while SWE-bench Verified is reported per successfully completed task.}
    \label{fig:charc_sens_workload}
\end{figure}

\begin{figure}[t]
    \centering
    \includegraphics[width=\linewidth]{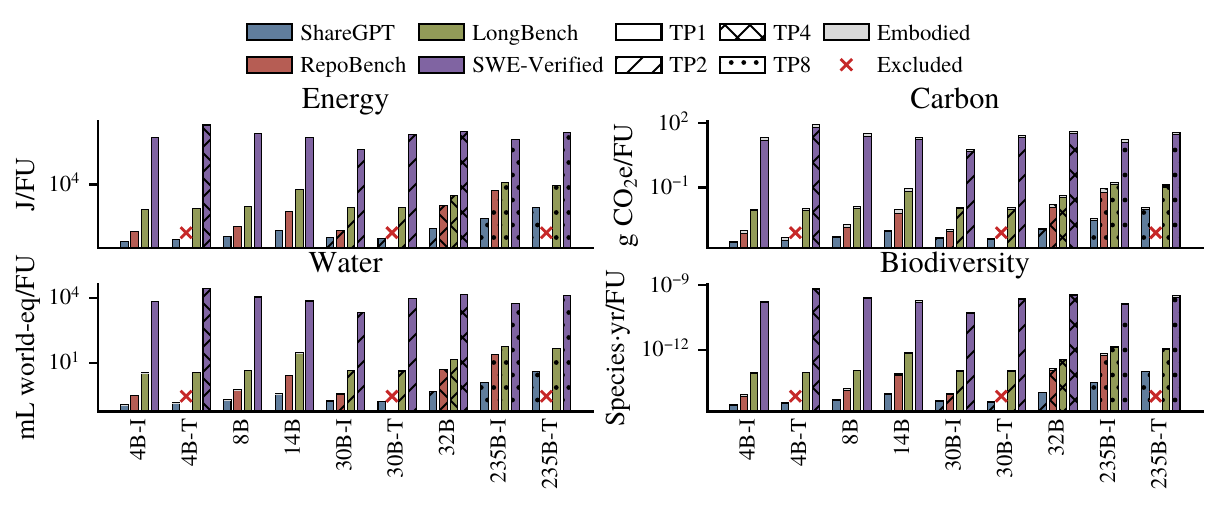}
    \caption{Impact variation across workloads for Qwen models on H100. Each model--workload pair uses its most energy-efficient tensor-parallel configuration. Non-agentic workloads are reported per request, while SWE-bench Verified is reported per successfully completed task. Crosses mark excluded configurations.}
    \label{fig:app_workload_qwen}
\end{figure}

\begin{figure}[t]
    \centering
    \includegraphics[width=\linewidth]{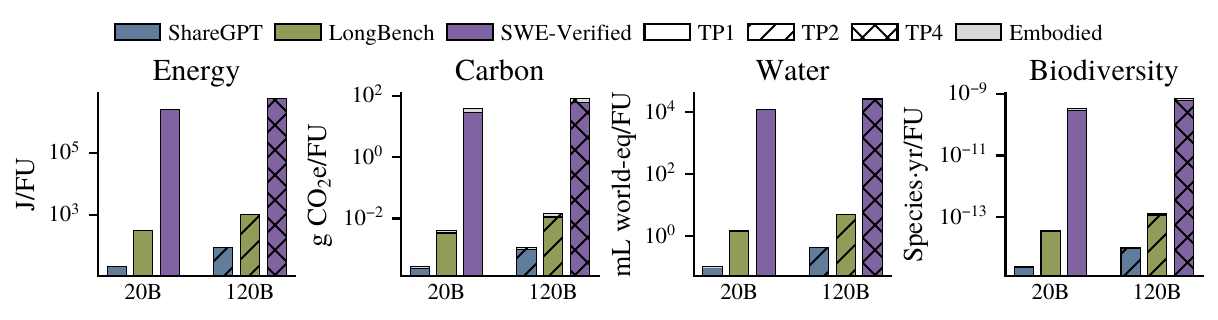}
    \caption{Impact variation across workloads for GPT-OSS models on H100. Each model--workload pair uses its most energy-efficient tensor-parallel configuration. Non-agentic workloads are reported per request, while SWE-bench Verified is reported per successfully completed task.}
    \label{fig:app_workload_gptoss}
\end{figure}

\begin{figure}[t]
    \centering
    \includegraphics[width=\linewidth]{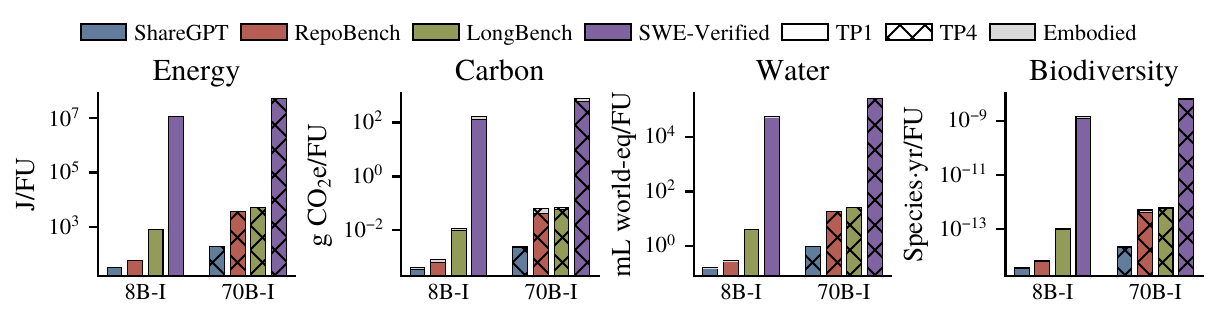}
    \caption{Impact variation across workloads for Llama models on H100. Each model--workload pair uses its most energy-efficient tensor-parallel configuration. Non-agentic workloads are reported per request, while SWE-bench Verified is reported per successfully completed task.}
    \label{fig:app_workload_llama}
\end{figure}

\noindent\textbf{Cross-workload variation.}
The preceding figures hold the workload fixed and vary the model. We next take the complementary view by fixing the model family and hardware and comparing workloads directly.
\Cref{fig:charc_sens_workload,fig:app_workload_qwen,fig:app_workload_gptoss,fig:app_workload_llama} show consistent workload-level patterns across Gemma, Qwen, GPT-OSS, and Llama. Among the non-agentic workloads, ShareGPT generally incurs the lowest per-request impact, while RepoBench and especially LongBench are higher because they process longer prompt and generation sequences. The magnitude of this increase depends on the model and its selected TP configuration, but energy, carbon, water, and biodiversity follow closely aligned trends within each matched workload comparison.

SWE-bench Verified is shown separately under a per-successful-task functional unit and therefore should not be compared numerically with the per-request bars. Its impact is substantially larger across all model families because an agentic task can require many model invocations and long execution, while low task success further increases impact per successful completion by amortizing
failed attempts over fewer successes.

\subsubsection{Additional Characterization of Spatiotemporal Impact Variations} \label{app:add_region_time}

\begin{figure}[t]
    \centering
    \includegraphics[width=\linewidth]{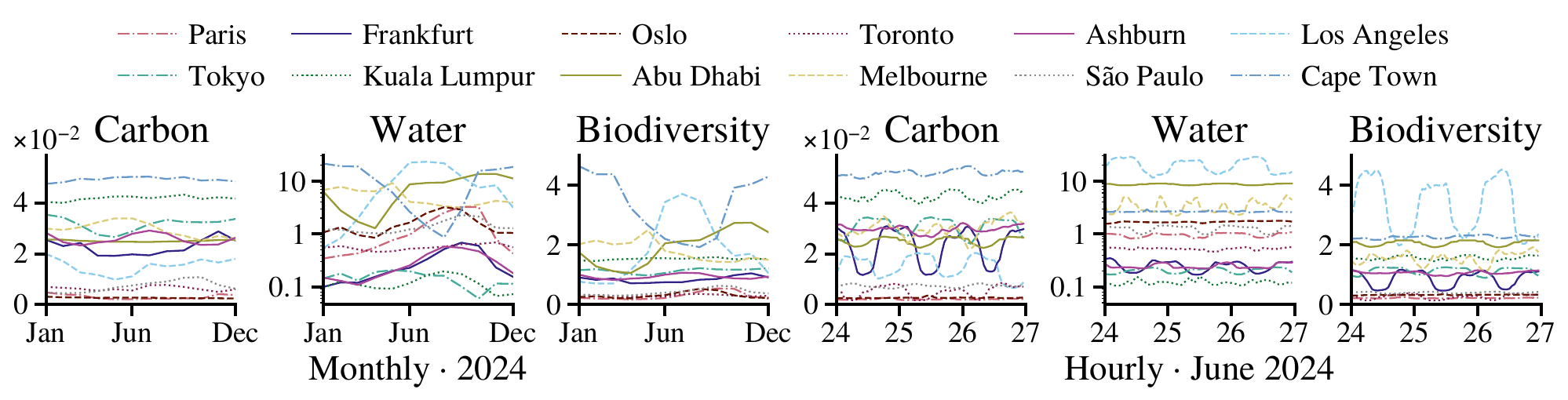}
    \caption{Temporal variation in per-request carbon, water, and biodiversity impact across selected deployment regions under the same serving setting as \Cref{fig:regional_comparison_simple}. Left: monthly averages in 2024; right: hourly variation during June 24--26, 2024. Carbon, water, and biodiversity are reported in g CO$_2$e/request, mL world-eq/request, and $10^{-13}$ species$\cdot$yr/request, respectively.}
    \label{fig:main_regional_time_sens}
\end{figure}

\begin{figure}[t]
    \centering
    \includegraphics[width=\linewidth]{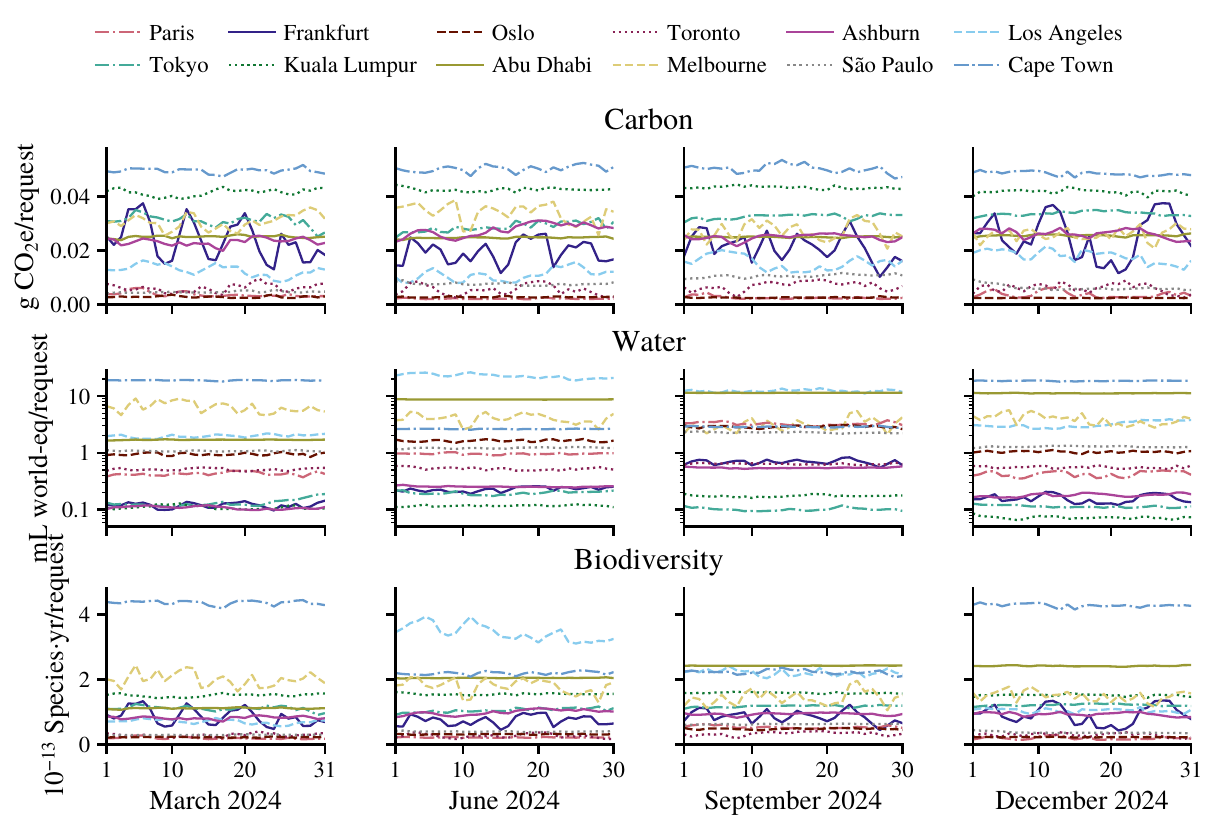}
    \caption{Daily variation in per-request carbon, water, and biodiversity impact across selected deployment regions in March, June, September, and December 2024 under the same serving setting as \Cref{fig:regional_comparison_simple}.}
    \label{fig:app_regional_impact_daily}
\end{figure}

\Cref{fig:main_regional_time_sens} provides the monthly and hourly results summarized in the main text. The ranking of regions changes over time and exhibit multiple crossovers, showing that annual-average preferences need not persist at finer temporal resolutions. The magnitude and timing of these changes differ across carbon, water, and biodiversity because each dimension depends on a different combination of regional environmental factors. \Cref{fig:app_regional_impact_daily} provides the complementary daily view for representative months across 2024 and shows that the same temporal variation is also visible at an intermediate timescale.

\begin{figure}[t]
    \centering
    \includegraphics[width=\linewidth]{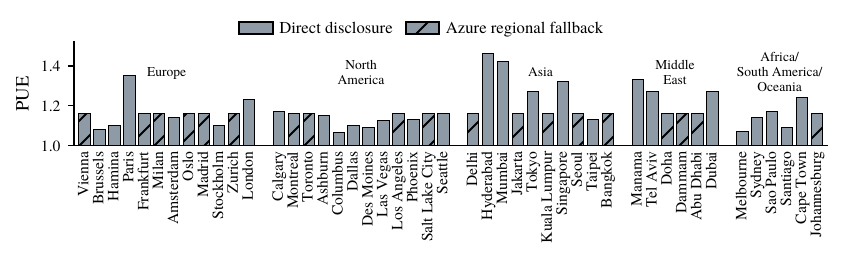}
    \caption{Annual-average PUE values for the modeled regions in \Cref{tab:modelled-deployment-regions}, based on public disclosures from \citet{aws-cloud-sustainability}, \citet{google-datacenter-pue}, and \citet{azure-datacenter-efficiency}.}
    \label{fig:app_regional_pue}
\end{figure}

\begin{figure}[t]
    \centering
    \includegraphics[width=\linewidth]{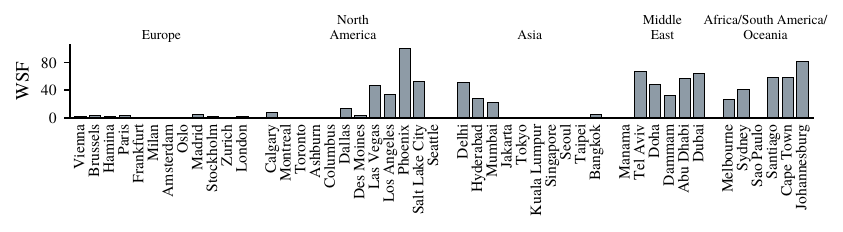}
    \caption{AWARE-2.0 water-scarcity factors for the modeled regions~\citep{seitfudem2025updated}.}
    \label{fig:app_annual_aware}
\end{figure}

\begin{figure}[t]
    \centering
    \includegraphics[width=\linewidth]{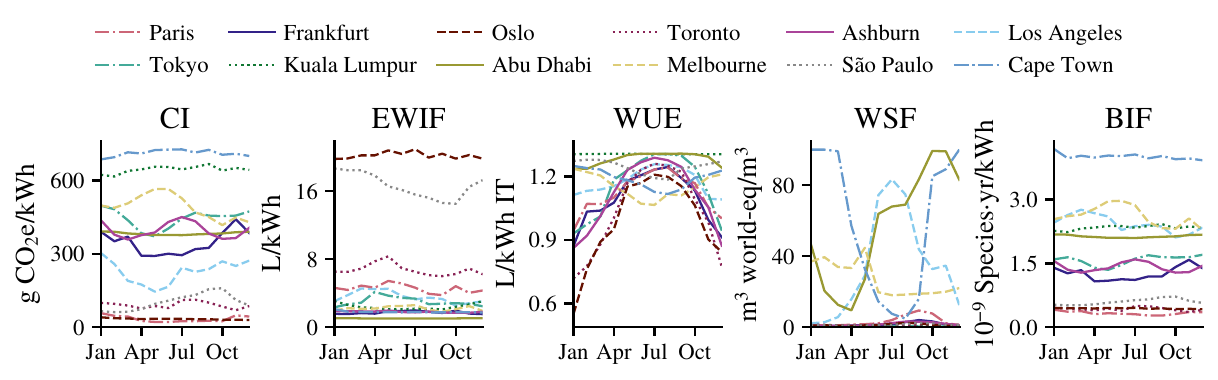}
    \caption{Monthly CI, EWIF, WUE, and $\mathrm{BIF}$ for selected modeled regions in 2024, together with the region-specific WSF used for water-scarcity adjustment. CI, EWIF, and $\mathrm{BIF}$ follow the observed electricity-grid mix, WUE reflects cooling conditions, while WSF captures regional water scarcity.}
    \label{fig:app_regional_alpha_monthly}
\end{figure}

\begin{figure}[t]
    \centering
    \includegraphics[width=\linewidth]{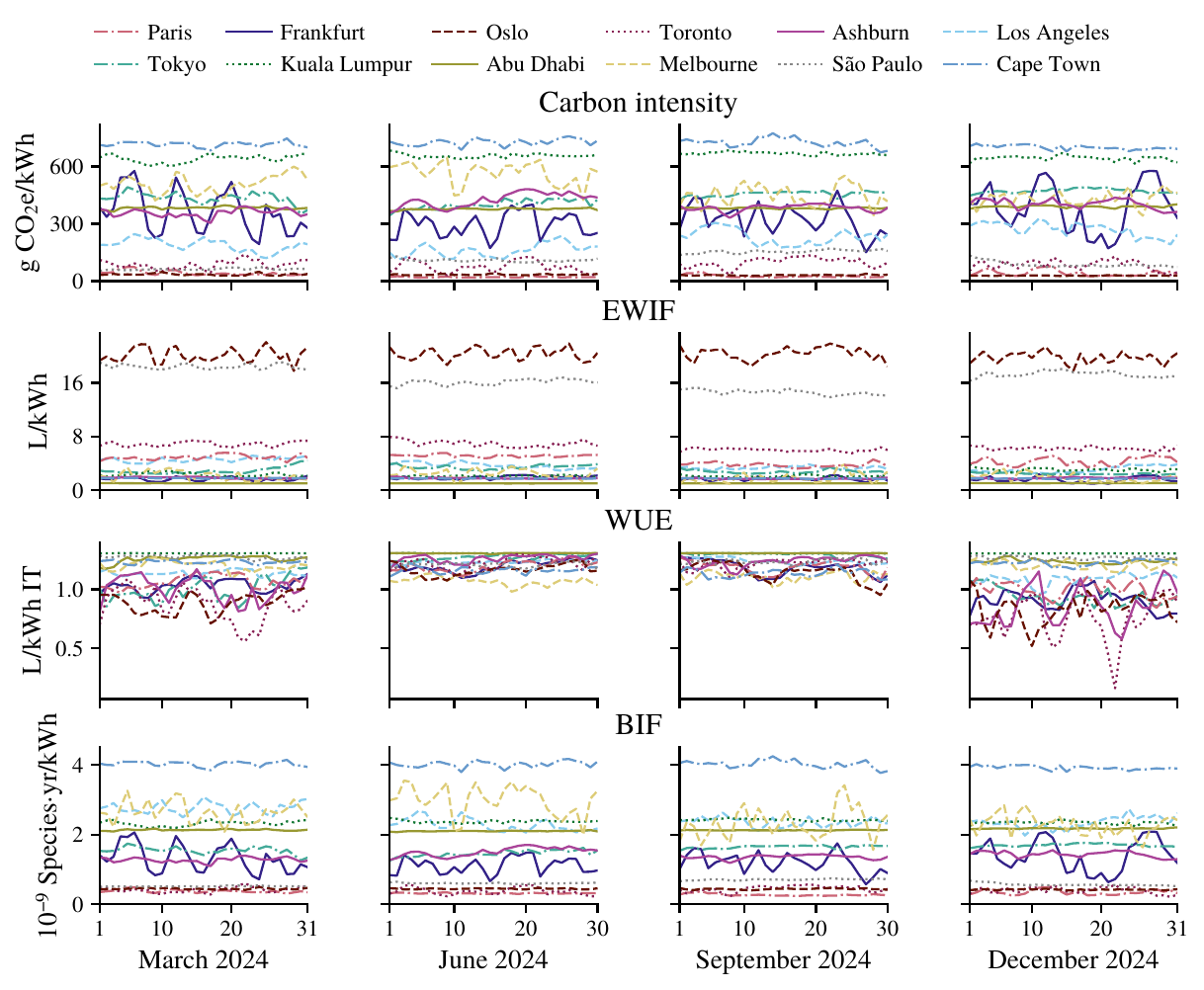}
    \caption{Daily CI, EWIF, WUE, and $\mathrm{BIF}$ for selected modeled regions in March, June, September, and December 2024.}
    \label{fig:app_regional_alpha_daily}
\end{figure}

\begin{figure}[t]
    \centering
    \includegraphics[width=\linewidth]{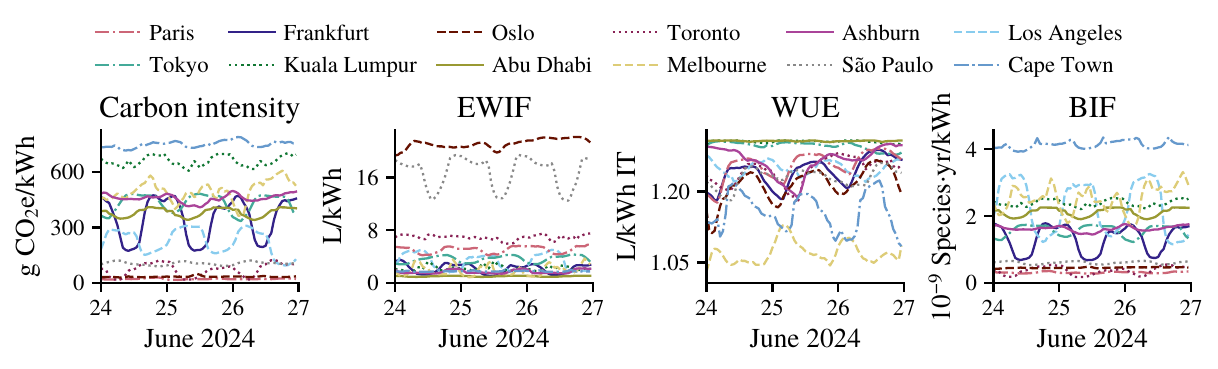}
    \caption{Hourly CI, EWIF, WUE, and $\mathrm{BIF}$ for selected modeled regions during June 24--26, 2024.}
    \label{fig:app_regional_alpha_hourly}
\end{figure}

The remaining figures expose the environmental data underlying these impact variations. \Cref{fig:app_regional_pue} reports the annual-average PUE values used for facility overhead, while \Cref{fig:app_annual_aware} reports the AWARE-2.0~\citep{seitfudem2025updated} water-scarcity characterization factors as the water stress factor (WSF). \Cref{fig:app_regional_alpha_monthly,fig:app_regional_alpha_daily,fig:app_regional_alpha_hourly} show the corresponding temporal behavior of grid carbon intensity (CI), electricity water-intensity factor (EWIF), direct water usage effectiveness (WUE), and grid biodiversity intensity $\mathrm{BIF}$ at monthly, daily, and hourly resolutions. Together, these data explain why the environmental dimensions exhibit different regional and temporal variation even when the serving workload and configuration are fixed.

\subsubsection{Sensitivity to Embodied Manufacturing Energy}
Our primary analysis defines the energy dimension as operational energy, consistent with \Cref{eq:energy_impact}. As a sensitivity analysis, we additionally account for energy consumed during device manufacturing and amortize it using the same allocation procedure as the other embodied impacts. \Cref{tab:app_emb_energy_contribution} shows the ranges of embodied energy contribution across the characterization and configuration-disagreement figures considered in the main text and appendix. Overall, embodied energy contribute to 1.35-6.93\% of lifecycle energy and do not change our conclusions on characterization results and cross-dimensional disagreements.

\begin{table*}[t]
    \centering
    \caption{Ranges of embodied energy contribution to lifecycle energy impact across the evaluated settings in the characterization and disagreement analyses.}
    \label{tab:app_emb_energy_contribution}

    \setlength{\tabcolsep}{5pt}
    \renewcommand{\arraystretch}{1.0}

    \begin{tabular}{@{}lll@{\hspace{2em}}lll@{}}
        \toprule
        Figure & Low (\%) & High (\%) &
        Figure & Low (\%) & High (\%) \\
        \midrule

        \Cref{fig:main_charac}
        & 1.35 & 3.72 &
        \Cref{fig:app_charac_longbench}
        & 2.18 & 3.90  \\

       \Cref{fig:regional_comparison_simple}
        & 2.23 & 2.64  & 
        \Cref{fig:charc_sens_workload}
        & 1.54 & 5.30  \\

        \Cref{fig:main_optimal_flip}
        & 1.83 & 2.51 &
        \Cref{fig:app_workload_qwen} & 1.36 & 6.44 \\

        \Cref{fig:charc_sens_hardware}
        & 1.54 & 3.83 &
        \Cref{fig:app_workload_gptoss}
        & 1.43 & 4.12 \\

        \Cref{fig:app_charc_sens_hardware_llama}
        & 1.48 & 2.09  & 
        \Cref{fig:app_workload_llama} & 1.35 & 5.89  \\

        \Cref{fig:app_charc_sens_hardware_gptoss}
        & 1.43 & 2.23  &
        \Cref{fig:app_optimal_config_disagreement}
        & 3.83 & 6.93 \\

        \Cref{fig:app_charc_sens_hardware_qwen}
        & 1.36 & 3.72 &  \Cref{fig:app_model_rank_disagreement}
        & 1.43 & 4.30 \\

        \Cref{fig:app_charac_repobench}
        & 2.84 & 6.44  &  \Cref{fig:main_hardware_flip}
        & 1.92 & 4.30\\

        \bottomrule
    \end{tabular}
\end{table*}

\subsection{Additional Cross-Dimensional Configuration-Ranking Disagreement}\label{app:lifecycle_disagreement}

This section distinguishes three levels of lifecycle disagreement on configuration choice. \emph{Optimal-configuration disagreement} means that different dimensions select different minimum-impact configurations from the same feasible configuration set satisfying the specified service requirements. \emph{General configuration-ranking disagreement} means that dimensions order a broader candidate set differently, even when no common quality constraint is imposed. \emph{Pairwise ordering flips} refer to controlled comparisons in which two fixed configurations exchange order across dimensions. \Cref{prop:lifecycle_crossover} is a pairwise result: it predicts when one configuration overtakes another under a given dimension. Optimal-configuration disagreement arises only when such a pairwise crossover changes the minimum-impact feasible configuration.

\begin{figure}[t]
    \centering
    \includegraphics[width=\linewidth]{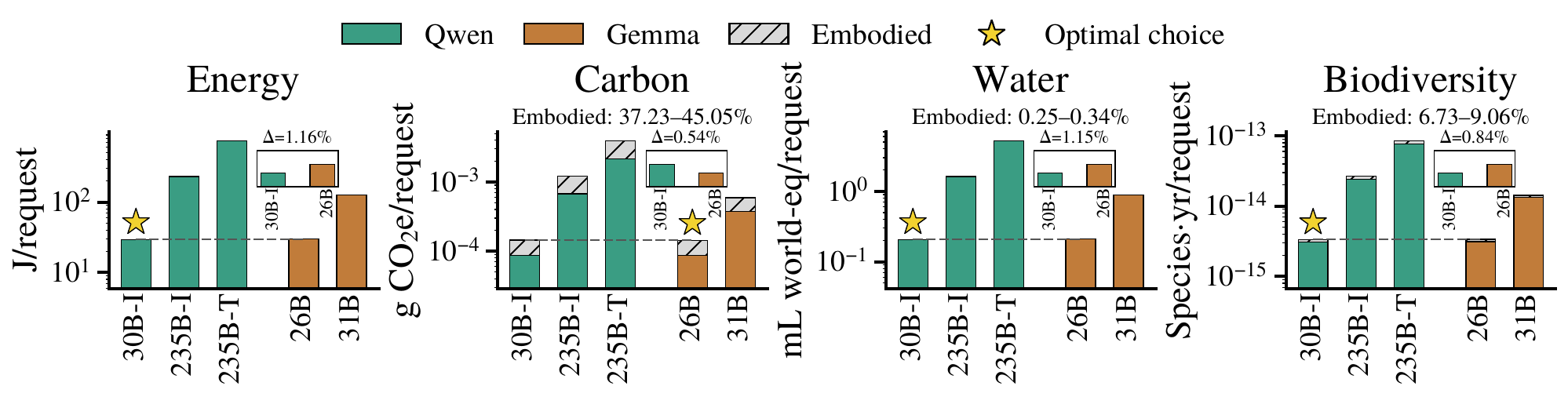}
    \caption{ShareGPT optimal configuration disagreement between dimensions}
    \label{fig:app_optimal_config_disagreement}
\end{figure}

\noindent \textbf{Optimal configuration disagreement.}
In addition to the RepoBench case in \Cref{fig:main_optimal_flip}, \Cref{fig:app_optimal_config_disagreement} shows a second quality-constrained optimal-configuration disagreement for ShareGPT. Under the $q_{\mathrm{chat}}\geq 86\%$ requirement in Norway, energy, water, and biodiversity select Qwen3 30B-Instruct on two H100s with TP2, whereas lifecycle carbon selects Gemma 4 26B on two H100s with TP2. As in the main-body example, the carbon-minimizing configuration under lifecycle accounting is not the lowest-energy feasible configuration because its embodied-carbon advantage is large enough to offset its additional operational-carbon penalty, thereby changing the minimum-impact feasible configuration.

\begin{figure}[t]
    \centering
    \includegraphics[width=\linewidth]{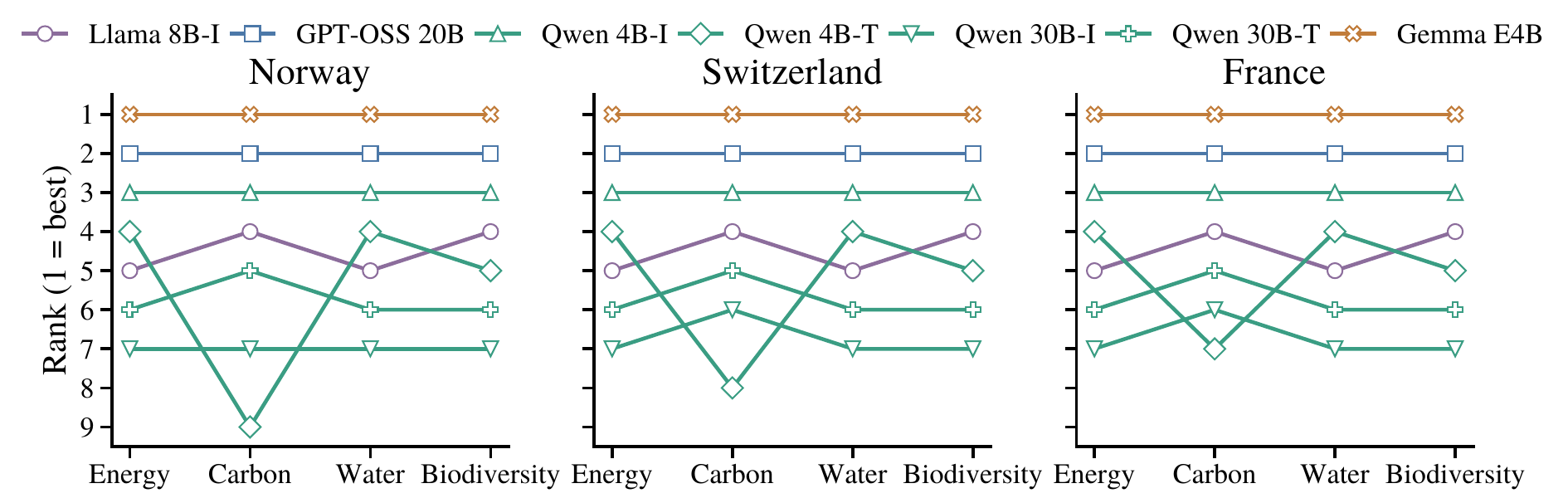}
    \caption{General model-ranking disagreement for ShareGPT across Norway, Switzerland, and France without imposing a quality constraint.}
    \label{fig:app_model_rank_disagreement}
\end{figure}
\noindent \textbf{General configuration-ranking disagreement.}
Optimal-configuration disagreement is the strongest decision-level outcome, but lifecycle effects can also alter the broader ordering of candidate models even without a specified quality threshold. \Cref{fig:app_model_rank_disagreement} illustrates this unconstrained ranking effect for ShareGPT across Norway, Switzerland, and France. 
Lifecycle carbon produces the largest ranking changes, particularly among the middle-ranked Qwen and Llama models, and these changes vary by region. Water largely preserves the configuration ranking induced by energy, while biodiversity introduces only minor additional swaps. Nevertheless, Gemma E4B remains the minimum-impact model across all four dimensions and all three regions, illustrating that substantial ranking disagreement does not necessarily produce optimal-configuration disagreement.

\begin{figure}[t]
    \centering
    \includegraphics[width=\linewidth]{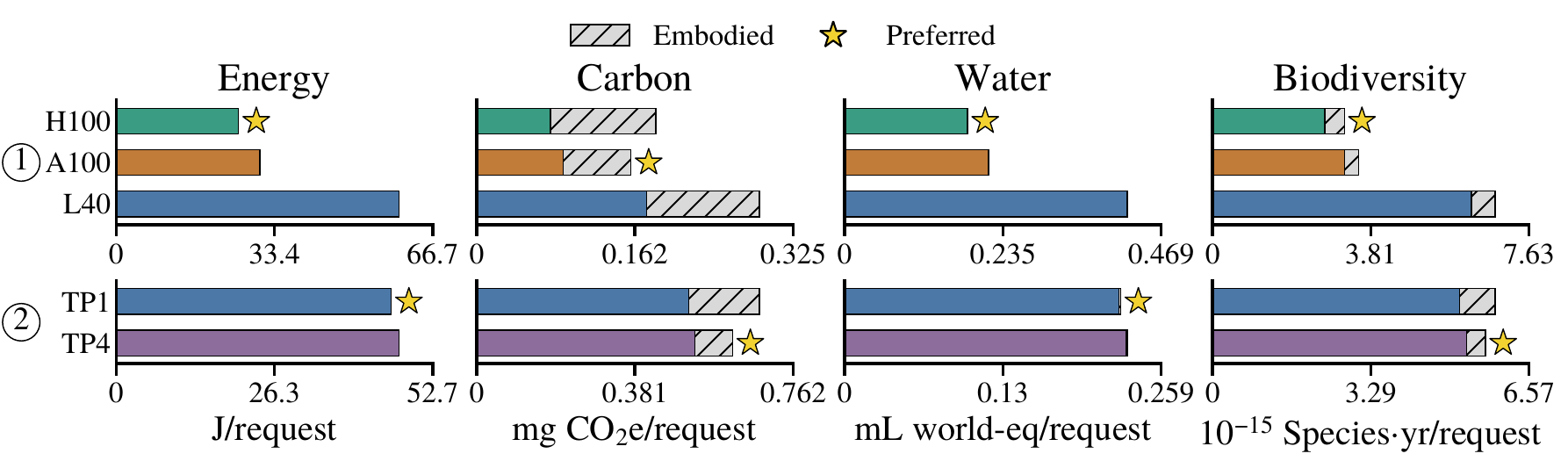}
    \caption{Lifecycle impact disagreement under controlled system choices. 
    \textcircled{1} Qwen3-4B-Thinking on ShareGPT in Norway, comparing H100, A100, and L40 at TP1. 
    \textcircled{2} Gemma 4 26B on ShareGPT in France, comparing H100 TP1 and TP4.
    Stars mark the preferred choice within each comparison.}
    \label{fig:main_hardware_flip}
\end{figure}
\noindent \textbf{Pairwise ordering flips under controlled hardware and parallelism choices.}
Disagreement can also arise without changing the model. In \Cref{fig:main_hardware_flip} \textcircled{1}, we fix Qwen3-4B-Thinking and TP1 and compare GPU platforms. H100 is preferred for energy, water, and biodiversity, whereas A100 is preferred for lifecycle carbon. In \Cref{fig:main_hardware_flip} \textcircled{2}, we fix Gemma 4 26B on H100 and compare only tensor-parallel settings. TP1 is preferred for energy and water, while TP4 is preferred for lifecycle carbon and biodiversity. These are controlled \emph{pairwise ordering flips}: they show that different dimensions can prefer different hardware or TP choices for the same model, but they do not by themselves imply that either configuration is the global optimum for the full candidate set.

\subsubsection{Numerical Validation of the Pairwise Crossover Boundary}
Across \Cref{fig:main_optimal_flip,fig:app_optimal_config_disagreement,fig:main_hardware_flip}, the common analytical object is the pairwise crossover condition in \Cref{eq:lifecycle_crossover}. For a lower-energy configuration $x_1$ and a higher-energy configuration $x_2$, we define the boundary multiple
\begin{equation}
\begin{aligned}
\Delta E_{\mathrm{IT}}
=
E_{\mathrm{IT}}(x_2,w)-E_{\mathrm{IT}}(x_1,w),
\quad
\Delta I_{m,\mathrm{emb}}
=
I_{m,\mathrm{emb}}&(x_1,w)-I_{m,\mathrm{emb}}(x_2,w), \\
\rho_m =
\frac{\Delta I_{m,\mathrm{emb}}/\Delta E_{\mathrm{IT}}}
{\alpha_m(r,t)},
\quad
\end{aligned}
\label{eq:boundary_multiple}
\end{equation}

A pairwise ordering flip under dimension $m$ occurs when $\rho_m>1$, i.e., when the embodied-impact advantage of $x_2$ exceeds its operational-impact disadvantage under deployment choice \((r,t)\). An optimal-configuration disagreement is a stronger outcome in which such a pairwise crossover changes the minimum-impact configuration among all quality- and SLO-feasible candidates. \Cref{tab:lifecycle_boundary_validation} evaluates this condition for representative optimal-configuration disagreement cases and controlled pairwise ordering flips.

\begin{table*}[t]
\centering
\caption{Numerical validation of the lifecycle crossover boundary.
Bold $\rho_m>1$ indicates a predicted lifecycle ranking reversal.}
\label{tab:lifecycle_boundary_validation}

\footnotesize
\setlength{\tabcolsep}{4.5pt}
\renewcommand{\arraystretch}{1.08}

\begin{tabular}{@{}llrrrrr@{}}
\toprule
\textbf{Workload / configuration pair}
& \textbf{Dimension}
& $\boldsymbol{\Delta E_{\mathrm{IT}}}$
& $\boldsymbol{\Delta I_{m,\mathrm{emb}}/}$
& $\boldsymbol{\alpha_m(r,t)}$
& $\boldsymbol{\rho_m}$
& \textbf{Critical lifetime} \\
 ($x_1\!\rightarrow\!x_2$) &
&
\textbf{(J/req.)}
& $\boldsymbol{\Delta E_{\mathrm{IT}}}$
 &
&
&
\textbf{(yr)} \\
\midrule

\makecell[l]{RepoBench (\Cref{fig:main_optimal_flip})\\Qwen3 4B-I $\rightarrow$ Qwen3 30B-I}
& Carbon & 9.422 & 5.342 & 2.924 & \textbf{1.827} & 10.96 \\

\makecell[l]{ShareGPT (\Cref{fig:app_optimal_config_disagreement}) \\ Qwen3 30B-I $\rightarrow$ Gemma 4 26B}
& Carbon & 0.341 & 5.201 & 2.924 & \textbf{1.778} & 10.67 \\

\midrule

\multirow{3}{*}{
    \makecell[l]{\Cref{fig:main_hardware_flip} \textcircled{1}\\
    H100 TP1 $\rightarrow$ A100 TP1}
}
& Carbon & 4.482 & 8.566 & 2.924 & \textbf{2.929} & 17.58 \\
& Water & 4.482 & 0.106 & 7.016 & 0.015 & 0.09 \\
& Biodiversity & 4.482 & 0.288 & 1.047 & 0.275 & 1.65 \\

\midrule

\multirow{3}{*}{
    \makecell[l]{\Cref{fig:main_hardware_flip} \textcircled{2}\\
    H100 TP1 $\rightarrow$ H100 TP4}
}
& Carbon & 1.248 & 62.466 & 11.143 & \textbf{5.606} & 33.64 \\
& Water & 1.248 & 0.622 & 4.904 & 0.127 & 0.76 \\
& Biodiversity & 1.248 & 2.719 & 1.120 & \textbf{2.429} & 14.57 \\

\bottomrule
\end{tabular}

\vspace{2pt}
\parbox{\textwidth}{\footnotesize
\textit{Units for $\Delta I_{m,\mathrm{emb}}/\Delta E_{\mathrm{IT}}$ and $\alpha_m$:}
Carbon: $10^{-6}$ g CO$_2$e/J;
Water: $10^{-6}$ L world-eq/J;
Biodiversity: $10^{-16}$ species$\cdot$yr/J.
Critical lifetime is the hardware amortization lifetime at $\rho_m=1$ with measured throughput fixed.
}
\end{table*}

The measured preferences agree with the analytical boundary. For the two quality-constrained optimal-choice cases (\Cref{fig:main_optimal_flip,fig:app_optimal_config_disagreement}), lifecycle carbon exceeds the crossover boundary by $1.78\times$ and $1.83\times$, respectively. The controlled comparisons reproduce the dimension-specific disagreements in \Cref{fig:main_hardware_flip}: for \textcircled{1}, only carbon crosses the boundary ($2.93\times$), whereas water and biodiversity remain well below it; for \textcircled{2}, carbon ($5.61\times$) and biodiversity ($2.43\times$) cross, while water does not. The critical-lifetime values provide an equivalent interpretation under our six-year amortization assumption: the corresponding pairwise preference persists while the assumed hardware lifetime remains below the listed threshold.

More generally, these results show why embodied share alone does not determine a decision change: the embodied difference must have the appropriate direction and be sufficiently large relative to both the operational-energy gap and the regional operational intensity. Intuitively, disagreement is easiest to trigger when the deployment choice has a low operational intensity for the dimension under consideration—for example, when carbon intensity is very low for the carbon dimension—because the operational penalty of a higher-energy configuration becomes small enough for embodied-impact differences to overturn the ranking.

\subsection{Supplementary Optimization Details}
\label{app:optimization_details}
This section provides the additional methodology and sensitivity analysis for the multidimensional routing optimization in the main text. We first describe how the mixed code and conversation workload is constructed from the Azure LLM Inference Dataset and paired with hourly environmental traces, followed by the synthetic regional-capacity model used to represent heterogeneous accelerator availability. We then give the full routing formulation, including the single-dimension optima and the minimax-regret objective used by \SYSTEM{}, and report implementation, baseline, solver, and runtime details. Finally, we evaluate sensitivity to workload composition, deployment-region scope, and regional-capacity realizations to test observed reduction in maximum normalized regret persists beyond the main experimental setting.
\subsubsection{Trace and Workload Construction}
\label{app:opt_trace}

We construct the routing workload from the Azure LLM Inference Dataset 2024. The code stream uses requests from May 10, 2024, and the conversation stream uses requests from May 12, 2024. We retain request arrival time and input/output token lengths, map each request to the corresponding measured H100 serving profile, and align the two streams by relative hour. The two streams are scaled to contribute equal offered GPU demand.

The optimization horizon contains 24 one-hour intervals. Requests with the same hour, traffic stream, serving profile, and token-length bins are aggregated into a weighted request group. Each group records the number of equivalent requests and may be divided across regions, corresponding to routing interchangeable requests in different proportions. This produces 1,628 request groups representing approximately 25.2 million request equivalents. Each traffic stream contributes 15.4 million GPU-seconds of offered demand.

We pair the demand trace with hourly environmental factors from June 24, 2024. The trace supplies the within-day demand pattern, while the environmental dataset supplies temporal variation in regional impact intensities. For workload sensitivity, we additionally construct a code-only instance by retaining only the code stream and applying the same preprocessing and capacity-generation procedure.

\subsubsection{Regional Capacity Model}
\label{app:opt_capacity}

We evaluate twelve deployment regions: Abu Dhabi, Melbourne, S\~ao Paulo, Toronto, Frankfurt, Paris, Tokyo, Kuala Lumpur, Oslo, Los Angeles, Northern Virginia, and Cape Town, as the regions studied in \Cref{fig:main_regional_time_sens}. Because regional accelerator inventories are not publicly available, we synthesize heterogeneous capacity.

For each capacity seed, we draw a lognormal weight for each region with log-space standard deviation $\sigma=0.5$ and normalize the weights to obtain fixed regional capacity shares. In each hour, total capacity of all regions is provisioned at $1.25\times$ the offered GPU demand. Regional allocations are converted to GPU counts, rounded upward to multiples of 32 H100 GPUs (one rack of 4 DGX H100 systems following NVIDIA’s H100 SuperPOD design~\citep{nvidia-dgx-superpod-h100}), and converted back to GPU-seconds. We evaluate seeds 0--19, with all routing policies using identical capacities for a given seed.

\subsubsection{Routing Optimization}
\label{app:opt_formulation}

Each workload uses its feasible configuration with the lowest measured IT energy. Let $y_{ir}$ denote the fraction of request group $i$ routed to region $r$. Each group must be fully assigned:
\begin{equation}
    \sum_r y_{ir}=1,
    \qquad
    0\leq y_{ir}\leq 1.
\end{equation}

Let $q_i$ denote the number of equivalent requests in group $i$, $g_i$ its GPU-seconds per request, and $t_i$ its arrival hour. Regional capacity is constrained by
\begin{equation}
    \sum_{i:t_i=t} q_i\, g_i\, y_{ir}
    \leq K_{rt},
    \qquad \forall r,t,
\end{equation}
where $K_{rt}$ is the available GPU-seconds in region $r$ during hour $t$. Requests are routed within their arrival hour; the experiment does not defer or drop demand. Thus, the time component of each deployment choice is fixed to the request group's arrival hour \(t_i\), and the optimization varies only the deployment region.

For each impact dimension $m\in\{E,C,W,B\}$, we first solve the corresponding single-dimension routing problem to obtain the feasible optimum $I_m^*$. Let \(I_m(y)\) denote the aggregate impact induced by routing assignment \(y\). \SYSTEM{} then solves
\begin{equation}
\begin{aligned}
    \min_{y,z}\quad & z \\
    \text{s.t.}\quad
    & I_m(y) \leq (1+z)I_m^*,
    && \forall m\in\{E,C,W,B\},
\end{aligned}
\label{eq:offline_prism_lp}
\end{equation}
together with the assignment and capacity constraints above. Since aggregated request groups are divisible, the main experiment is a linear program (LP). Representing individual requests with indivisible assignments gives the corresponding mixed-integer formulation (MILP).

All routing policies return assignments to the same impact evaluator, which computes energy, carbon, water, and biodiversity using the accounting model in \Cref{sec:sustainability_dimensions}.

\subsubsection{Implementation details}
\label{app:opt_implementaion}

\noindent \textbf{Experimental Configurations.} Unless otherwise stated, experiments use a 24-hour horizon with one-hour routing intervals and the twelve-region deployment set. Code and conversation traffic contribute equal offered GPU demand. Total regional capacity in each hour is provisioned at $1.25\times$ offered demand, with regional shares drawn from a lognormal distribution with $\sigma=0.5$ and held fixed over time. Capacity is rounded upward to 32-H100 units, and results are reported across capacity seeds 0--19. Each workload uses its minimum-IT-energy feasible computing configuration \(x_E^*(w)\). All demand must be served in its arrival hour; requests are neither dropped nor deferred.

\noindent \textbf{Baselines.} The single-dimension baselines minimize energy, carbon, water, or biodiversity under the same assignment and capacity constraints. The load-balancing baseline uses regional capacity without environmental information.
Our offline WaterWise~\citep{jiang2025waterwise} adaptation preserves its joint carbon--water objective. Carbon and water are normalized across eligible regions and weighted equally. We evaluate its assignments using the same carbon and stress-adjusted water accounting as the other policies. Since the routing experiment contains neither request deferral nor request-origin information, only the spatial routing component is used.

\noindent\textbf{Solver and Runtime.}
We implement the optimization in Python 3.12 using SciPy 1.17.1~\citep{virtanen2020scipy} with the HiGHS linear-optimization backend~\citep{huangfu2018parallelizing}. The twelve-region instance contains 1,628 request-group assignment constraints, 288 region-hour capacity constraints, and 19,536 routing variables. \SYSTEM{} adds one maximum-regret variable and four regret constraints. On one AMD EPYC 7443 CPU, the complete set of optimization policies for the twelve-region instance requires less than 4\,s end-to-end, including constraint construction and impact evaluation. For \SYSTEM{} alone, the optimization solve takes below 0.3\,s in our measurements.

\subsubsection{Additional Sensitivity Results}
\label{app:opt_sensi}

We additionally evaluate on the code-only instance defined in \Cref{app:opt_trace}. \SYSTEM{} obtains 37.0\% worst-case regret, compared with 74.5\% for WaterWise, 66.9\% for water-only routing, 384.4\% for biodiversity-only routing, 466.0\% for load balancing, 708.1\% for energy-only routing, and 838.6\% for carbon-only routing. The result is consistent with the mixed-workload experiment: balancing all four dimensions substantially reduces the maximum normalized regret.

We also examine the influence of regional scope and capacity. We repeat the experiment across 20 capacity seeds under three regional scopes. The \emph{original six-region} set contains Frankfurt, Los Angeles, Tokyo, Kuala Lumpur, Abu Dhabi, and Melbourne, matching the representative locations in \Cref{fig:regional_comparison_simple}. The \emph{all-twelve-region} set additionally includes Paris, Toronto, Northern Virginia, Oslo, Cape Town, and S\~ao Paulo and is used for the main optimization result. The \emph{concentrated six-region} set contains Frankfurt, Paris, Los Angeles, Northern Virginia, Toronto, and Tokyo. It provides a less geographically dispersed deployment space concentrated in Europe, North America, and East Asia, allowing us to test whether the result depends on the broader geographic and environmental diversity of the twelve-region set.

\begin{table}[t]
    \centering
    \caption{Sensitivity of worst-case regret to geographic scope and capacity seed. Values report median [range] across 20 seeds.}
    \label{tab:opt_region_sensitivity}
    \begin{tabular}{lccc}
        \toprule
        \textbf{Region set}
        & \textbf{WaterWise}
        & \textbf{\SYSTEM{}}
        & \textbf{Median reduction} \\
        \midrule
        Original 6
        & 43.5\% [28.8, 111.2]
        & 37.2\% [25.5, 53.5]
        & 19.1\% \\
        All 12
        & 157.8\% [55.1, 212.3]
        & 75.1\% [41.4, 104.0]
        & 50.2\% \\
        Concentrated 6
        & 101.7\% [57.7, 208.5]
        & 68.1\% [48.5, 107.4]
        & 37.9\% \\
        \bottomrule
    \end{tabular}
\end{table}

Across all three region sets and 60 capacity instances, \SYSTEM{} achieves lower worst-case regret than WaterWise. The benefit is largest for the twelve-region design space, where greater environmental heterogeneity creates larger trade-offs among the four impact dimensions.

\subsection{Extension of Optimization} \label{app:opt_ext}
This section extends the main optimization study beyond its offline, fixed-computing-configuration setting in two directions. First, we evaluate \SYSTEM{} under rolling-horizon regional routing, where decisions are repeatedly re-optimized using limited and potentially noisy forecasts of future environmental conditions. Second, we consider an agent-heavy workload and jointly optimize computing configuration and deployment region while introducing agent completion time as an additional objective alongside energy, carbon, water, and biodiversity. These extensions test whether the multidimensional optimization remains effective under sequential decision-making and whether its formulation can accommodate configuration choice and performance trade-offs beyond regional routing alone.
\subsubsection{Rolling-Horizon Regional Routing}
\label{app:opt_online}

We extend \SYSTEM{} to rolling-horizon routing using the same 24-hour workload, twelve regions, fixed computing configurations, and seed-0 capacities as the offline experiment. At the beginning of each hour, the scheduler observes the current state and optimizes over a forecast horizon
$H\in\{1,6,12,24\}$ hours. It executes only the current-hour allocation and re-optimizes at the next hour; requests remain in their arrival hour.

At each optimization step, projected regret combines impacts already realized with forecast impacts over the remaining window and compares them against the corresponding cumulative single-dimension optima. We evaluate both oracle forecasts and noisy environmental forecasts. For the latter, future CI, EWIF, WUE, and BIF values are independently perturbed by mean-one lognormal noise with coefficient of variation $0.20$; the current hour is observed exactly. Demand and capacity are assumed known. Full-day offline \SYSTEM{} provides the reference optimum, and we use the same WaterWise-spatial baseline as in \Cref{app:opt_implementaion}.

\begin{table}[t]
    \centering
    \caption{Rolling-horizon \SYSTEM{} under oracle and noisy environmental forecasts. Gap is the relative increase in maximum regret over full-day offline \SYSTEM{}.}
    \label{tab:opt_rolling_horizon}
    \begin{tabular}{lrrrr}
        \toprule
        & \multicolumn{2}{c}{\textbf{Oracle forecast}}
        & \multicolumn{2}{c}{\textbf{20\% forecast error}} \\
        \cmidrule(lr){2-3}\cmidrule(lr){4-5}
        \textbf{Horizon}
        & \textbf{Max. regret}
        & \textbf{Gap}
        & \textbf{Max. regret}
        & \textbf{Gap} \\
        \midrule
        1 h  & 88.014\% & 0.697\% & 88.014\% & 0.697\% \\
        6 h  & 87.482\% & 0.088\% & 87.532\% & 0.146\% \\
        12 h & 87.415\% & 0.012\% & 87.517\% & 0.129\% \\
        24 h & 87.405\% & $<0.001$\% & 87.525\% & 0.138\% \\
        \midrule
        Offline \SYSTEM{}
             & 87.400\% & -- & -- & -- \\
        WaterWise-spatial
             & 188.240\% & 115.36\% & -- & -- \\
        \bottomrule
    \end{tabular}
\end{table}

As shown in \Cref{tab:opt_rolling_horizon}, even a one-hour oracle horizon remains within 0.70\% of the offline optimum in maximum regret. The gap falls to 0.088\% at six hours and 0.012\% at twelve hours, while the 24-hour horizon reproduces the offline solution to numerical precision. Environmental forecast error has little effect: for horizons containing unobserved future hours, 20\% factor noise leaves the gap below 0.15\%. In comparison, WaterWise-spatial incurs 188.2\% maximum regret. Thus, the multidimensional trade-off obtained offline is largely preserved under sequential routing and remains stable to moderate environmental forecast error.

Across the full 24-hour replay, rolling-horizon optimization requires 0.60--3.21\,s of solver time and 12.2--19.5\,s end-to-end across the evaluated horizons on the AMD EPYC 7443 system described in \Cref{app:opt_implementaion}.

\subsubsection{Joint Optimization with Agentic Workload}
\label{app:opt_agentic}

We extend the routing experiment to an agent-heavy workload and jointly optimize computing configuration and deployment region. To represent a 2026-style serving mix in which repeated agentic interactions dominate compute demand, the 24-hour workload consists of 60\% agentic coding, 30\% conversation, 5\% long-output summarization, and 5\% code completion by offered GPU-seconds. Conversation follows the Azure conversation trace with ShareGPT profiles, summarization uses the LongBench long-output workload, and code completion uses RepoBench. Agentic arrivals follow the hourly shape of the Azure code trace and use SWE-Bench Verified. Defining the mix by GPU demand rather than request count avoids treating a short inference request and a multi-step agentic task as equivalent units of load. We use the same twelve regions, hourly environmental factors, and capacity model as the offline routing experiment.

For agentic coding, the functional unit remains a successfully completed SWE-Bench Verified task. Each arrival can be assigned to any of 16 measured computing configurations, without assuming that the scheduler knows which model will solve an individual task. For configuration $x$, expected IT energy and GPU demand per successful task are obtained by dividing the full-suite benchmarking totals by its number of successful tasks, while $\tau_x$ is the mean completion time among successful tasks. The other workloads retain their fixed measured configurations.

\textbf{Completion time is introduced as an additional objective} because a hard timeout does not distinguish between otherwise feasible agent executions that finish substantially earlier or later. Let $y_{ir}^x$ be the fraction of request group $i$ assigned to configuration $x$ and region $r$, and let $\omega_i$ denote its weight. Each request group is fully assigned across configuration--region pairs:
\begin{equation}
    \sum_{x\in\mathcal X}\sum_{r\in\mathcal R} y_{ir}^{(x)} = 1,
\qquad
y_{ir}^{(x)} \ge 0,
\qquad \forall i.
\end{equation}
For the set of agentic groups $\mathcal{A}$, we define mean agentic completion-time objective as
\begin{equation}
    L(s)
    =
    \frac{1}{\Omega_{\mathcal A}}
    \sum_{i\in\mathcal A}\sum_{x,r}
    \omega_i \,y_{ir}^{(x)}\,\tau_x,
    \qquad
    \Omega_{\mathcal A}=\sum_{i\in\mathcal A}\omega_i .
\end{equation}
Let $L^*$ be the minimum feasible completion time under the same assignment and capacity constraints. Completion-time regret is
\begin{equation}
    R_L(s)=\frac{L(s)-L^*}{L^*}.
\end{equation}
For each environmental dimension $m\in\{E,C,W,B\}$, we similarly obtain its independent optimum $I_m^*$ and define
\begin{equation}
    R_m(s)=\frac{I_m(s)-I_m^*}{I_m^*}.
\end{equation}
Environmental \SYSTEM{} minimizes
$\max\{R_E,R_C,R_W,R_B\}$, whereas the five-objective formulation jointly minimizes
\begin{equation}
    \min_s
    \max\{R_E(s),R_C(s),R_W(s),R_B(s),R_L(s)\}.
    \label{eq:agentic_five_dimensional_prism}
\end{equation}

We compare these two formulations with energy-only and completion-time-only optimization and with two WaterWise adaptations. WaterWise-spatial first fixes each workload to its minimum expected-energy configuration and optimizes regional carbon and water, while WaterWise-joint allows the same carbon--water objective to choose both configuration and region.

\begin{figure}[t]
    \centering
    \includegraphics[width=\linewidth]{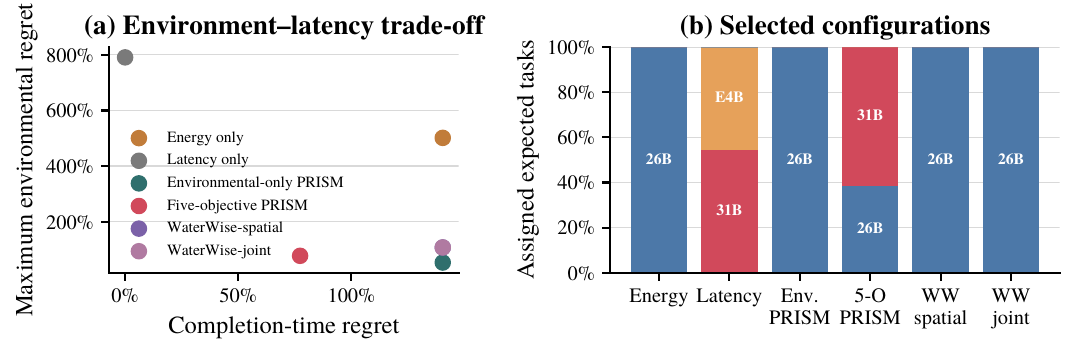}
    \caption{Joint computing-configuration and regional-routing optimization under the agent-heavy workload. (a) Maximum environmental regret versus agentic completion-time regret. (b) Configuration shares selected for the agentic workload.}
    \label{fig:agentic_joint_optimization}
\end{figure}

\Cref{fig:agentic_joint_optimization} shows that adding completion time changes both the trade-off and the selected configuration. Environmental-only \SYSTEM{} achieves 53.1\% maximum environmental regret but incurs 140.4\% completion-time regret. Five-objective \SYSTEM{} instead limits both to 77.3\%. Carbon, stress-adjusted water, and completion-time regrets are binding at the minimax solution, while energy and biodiversity regrets remain lower at 18.0\% and 42.1\%, respectively. By comparison, energy-only and completion-time-only optimization incur maximum environmental regrets of 501.4\% and 790.8\%, while WaterWise-spatial and WaterWise-joint obtain 107.4\% and 107.2\% environmental regret with 140.4\% completion-time regret.

The performance objective also changes configuration selection. Environmental \SYSTEM{} and both WaterWise variants assign all agentic work to Gemma 4 26B. Five-objective \SYSTEM{} instead assigns 38.4\% to Gemma 4 26B and 61.6\% to the faster Gemma 4 31B, reducing completion time until its regret reaches the environmental minimax boundary. Completion-time-only optimization shifts further toward faster configurations, assigning 54.4\% to Gemma 4 31B and 45.5\% to Gemma 4 E4B. Thus, once agent completion time is treated as an objective rather than only a feasibility constraint, computing configuration and regional routing should therefore be optimized jointly.

The resulting continuous linear programs contain 1,247 weighted request groups and 19,284 joint configuration--region assignment variables. Using the same SciPy/HiGHS implementation as \Cref{app:opt_implementaion}, individual solves require 0.05--0.16\,s in this experiment.

\end{document}